\documentclass[12pt]{article}

\usepackage{amsmath,amssymb,epsfig,subfigure,color,soul,fleqn,rotating}
\usepackage{ulem}
\usepackage{cite}
\usepackage[colorlinks,citecolor=blue,urlcolor=blue,linkcolor=blue]{hyperref}

\newcommand{\be}{\begin{eqnarray}}
\newcommand{\ee}{\end{eqnarray}}
\newcommand{\nee}{\nonumber\end{eqnarray}}

\newcommand{\msq}[1] {m_{\sq_{#1}}}

\newcommand{\mch}[1] {m_{\ti \x^+_{#1}}}
\newcommand{\mnt}[1] {m_{\ti \x^0_{#1}}}

\newcommand{\msg}    {m_{\ti g}}
\newcommand{\msu}[1] {m_{\ti u_{#1}}}
\newcommand{\msd}[1] {m_{\ti d_{#1}}}

\def\fb              {${\rm fb}^{-1}$}
\def\gev             {{\rm GeV}}

\newcommand{\gsim}{\;\raisebox{-0.9ex}
           {$\textstyle\stackrel{\textstyle >}{\sim}$}\;}

\def\al               {\alpha}
\def\be               {\beta}
\def\d               {\delta}

\def\x               {\chi}
\def\ti              {\tilde}
\def\sq              {\ti q}

\def\ch              {\ti \x^\pm}
\def\chp             {\ti \x^+}

\def\nt              {\ti \x^0}
\def\sg              {\ti g}

\def\bart              {\bar{t}}
\def\barc             {\bar{c}}
\def\su                {\ti{u}}
\def\sto                  {\ti{t}}
\def \sca                 {\ti{c}}
\def\sd                {\ti{d}}

\def\sbo                 {\ti{b}}
\def\dll            {\d^{LL}_{23}}
\def\durr            {\d^{uRR}_{23}}
\def\durl            {\d^{uRL}_{23}}
\def\dulr            {\d^{uLR}_{23}}
\def\ddrr            {\d^{dRR}_{23}}

\newcommand{\thw}{\theta_{\textit{\tiny{W}}}}
\newcommand{\etmiss}{{E_T \hspace{-4.5mm}/} \hspace{2.5mm}}

\newcommand{\AddrVienna}{
\it Universit\"at Wien, Fakult\"at f\"ur Physik,
A-1090 Vienna, Austria \\}

\newcommand{\AddrGAKUGEI}{%
 \it Department of Physics, Tokyo Gakugei University, Koganei,
Tokyo 184-8501, Japan\\}

\newcommand{\AddrHEPHY}{%
 \it Institut f\"ur Hochenergiephysik der \"Osterreichischen Akademie
der Wissenschaften, A-1050 Vienna, Austria\\}

\newcommand{\AddrWuerzburg}{%
 \it Institut f\"ur Theoretische Physik und Astrophysik, Universit\"at W\"urzburg,
D-97074 W\"urzburg, Germany\\}

\newcommand{\AddrLAPTh}{
\it LAPTh, Universit\'e de Savoie, CNRS, 9 Chemin de Bellevue, B.P. 110,
F-74941 Annecy-le-Vieux, France\\}

\title{Flavour violating bosonic squark decays at LHC}

\author{A. Bartl${}^{1}$, H. Eberl${}^2$, E. Ginina${}^{1}$, B. Herrmann${}^{3}$, K.~Hidaka${}^4$,\\
W.~Majerotto${}^2$ and W. Porod${}^{5}$}
\date{
\small  $^1$ \AddrVienna
        $^2$ \AddrHEPHY
        $^3$ \AddrLAPTh
        $^4$ \AddrGAKUGEI
        $^5$ \AddrWuerzburg}

\definecolor{darkgreen}{rgb}{0,.5,0}

\begin{document}

\begin{flushright}
LAPTH-050/12\\
UWThPh-2012-35\\
HEPHY-PUB 923/12
\end{flushright}
\begingroup
\let\newpage\relax
\maketitle
\endgroup

\maketitle
\thispagestyle{empty}

\begin{abstract}
We study quark flavour violation (QFV) in the squark sector of the Minimal Supersymmetric Standard Model (MSSM). 
We assume mixing between the second and the third squark generations, i.e. $\sca_{R}-\sto_{L, R}$ mixing. We focus on QFV 
effects in bosonic squark decays, in particular 
on the decay into the lightest Higgs boson $h^0$, $\su_2 \to \su_1 h^0$, where $\su_{1,2}$ are the lightest up-type squarks.
We show that the branching ratio of this QFV decay can be quite large (up to 50 $\%$) due to large QFV trilinear couplings,
and large $\sca_{R}-\sto_{L, R}$  and $\sto_{L}-\sto_{R}$ mixing, despite 
the strong constraints on QFV from B meson data.
This can result in characteristic QFV final states with significant rates at LHC (14 TeV), 
such as $pp \to \sg \sg X \to t + h^0 + 3\,$ jets$~+ \etmiss + X$ and 
$pp \to \sg \sg X \to t t~({\rm or}~\bar{t} \bar{t}) + h^0 + 2\,$ jets$~+ \etmiss + X$.
The QFV  bosonic squark decays can have an influence on the squark and gluino searches at LHC.

\vspace*{2cm}



\end{abstract}

\clearpage

\section{Introduction}
\label{sec:intro}

In most searches for supersymmetric (SUSY) particles at the LHC, the analyses have been performed within simplified SUSY models.
However, SUSY extensions of the Standard Model (SM) can have a richer structure.
In principle, mixing between the different squark generations is  possible in the  Minimal Supersymmetric Standard Model
(MSSM).
This can lead to quark flavour violating (QFV) effects, in addition to those induced by the Cabibbo-Kobayashi-Maskawa 
(CKM) matrix~\cite{Buras:2000dm, Ambrosio:2002ex, Kagan:2009bn}. 
The mixing structure of the squark sector may be completely uncorrelated to the CKM matrix. Therefore, a detailed study of the 
consequences of general squark mixing
is highly appropriate.
Mixing between the $1^{\rm{st}}$ and the $2^{\rm{nd}}$ squark generations is strongly supressed by K physics data~\cite{Beringer:1900zz}.
Therefore, in this paper we assume mixing between
the $2^{\rm{nd}}$ and the $3^{\rm{rd}}$ squark generations, respecting the constraints from B physics. 
Although these constraints are quite severe, they allow nevertheless substantial QFV effects. 

In the MSSM, the mixing of the $2^{\rm{nd}}$ and the $3^{\rm{rd}}$ squark generations was theoretically studied for squark and 
gluino production and their decays at the LHC in the context
of Minimal Flavour Violation (MFV) \cite{Hiller:2008wp,Hiller:2009ii,Muhlleitner:2011ww}
as well as for general flavour mixing
\cite{Bozzi:2007me,Fuks:2008ab,Bartl:2010du,%
Bruhnke:2010rh, Hurth:2009ke,Bartl:2009au, Bartl:2011wq, Fuks:2011dg, Nomura:2007ap}. As shown in these papers the effects of QFV can be large.
For example, in the case of mixing between scalar top and scalar charm, we can expect a large branching ratio (up to 40 $\%$) of 
the QFV decay of the gluino, $\sg \to c \bart (\barc t) \nt_1$~\cite{Bartl:2009au}. This is due to the fact that the lightest up--squark 
mass eigenstates $\su_{1,2}$ 
are mainly mixtures of $\sto_R$ and $\sca_R$.
Hence, $\su_1$ and $\su_2$ can both decay into $c \nt_1~{\it and}~t \nt_1$. 

In addition to the fermionic decays of squarks there are bosonic decays, 
$\sq_i \to \sq_j+Z^0, h^0, H^0, A^0$ and $\sq_i \to \sq_j'+ W^\pm, H^\pm$, if kinematically allowed. In the quark 
flavour conserving (QFC) case, the most interesting decays are
$\sto_2 \to \sto_1+Z^0, h^0, H^0, A^0$; $\sbo_2 \to \sbo_1+Z^0, h^0, H^0,$ $A^0$; $\sbo_2 \to \sto_1+W^-, H^-$~\cite{Bartl:1994bu,Bartl:1998xk}. 
The QFV bosonic decays were recently considered in \cite{Bruhnke:2010rh}. 
There the characteristic differences to the MFV case were worked out. A non--minimal flavour 
structure in the squark sector can change the entire squark decay pattern quite drastically, because many more transitions are possible. 

In the present paper, we study the bosonic decays of the up-type squarks, $\su_2 \to \su_1 h^0/Z^0,~ \su_3 \to \su_{1,2} h^0/Z^0$, in the MSSM. 
Motivated by the recently observed signal of a Higgs boson at LHC, we are particularly interested in the bosonic QFV squark decays into the lightest Higgs boson, 
$\su_{2} \to \su_1 h^0$. These decays offer the best possibility of determining the trilinear couplings $\sq_i-\sq_j-h^0$ 
entering the soft-SUSY-breaking Lagrangian. 
Another possibility would be to study the 3-body production $pp \to \sq_i \sq_j h^0$ as discussed for example in \cite{Djouadi:1999dg}
for the QFC case.
As the Higgs boson 
couples dominantly to the $\sq_L$--$\sq_R$ combination, 
one gets information from the decays $\sq_i \to \sq_j + h^0$ 
on the flavour structure of the left--right (LR) terms in the squark mass matrix. 
We study the mixing between the $2^{\rm{nd}}$ and the $3^{\rm{rd}}$ generation of up--type squarks, i.e. 
$\sca_{R}-\sto_{L,R}$ mixing.
There are strong constraints on this mixing from B physics (see also~\cite{Cao:2006xb}), Higgs boson searches and SUSY particle searches 
(see Appendix ~\ref{sec:exp_constr}).
We take into account all these constraints in our analysis. 
The QFV bosonic squark decays mentioned above have not been
explicitly 
searched for at LHC so far.
But these decays may show up at the higher energy run with $\sqrt{s}$ = 14~TeV at the LHC. We will
work out the most important QFV signatures of these bosonic decays.
 
The paper is organized as follows: In Section~\ref{sec:sq.matrix} we 
shortly give the definitions of the QFV squark mixing parameters. In Section~\ref{sec:bos_sq_decays} 
we discuss the QFV bosonic decays of up-type squarks in detail in a definite scenario accesible at LHC.
We also consider two further scenarios, one GUT inspired and another one, where the bosonic 
decays of $\su_2$ dominate over the fermionic decays. Section~\ref{sec:signatures} contains a discussion of various QFV final states to be expected at LHC with $\sqrt{s}=14$~TeV. 
In Section~\ref{sec:conclude} we give a summary. In the Appendices we 
show explicitly the part of the interaction Lagrangian which is most relevant for this study and
summarize the experimental and theoretical constraints on the MSSM parameters, especially those on the QFV parameters, mainly 
from B physics.

%
%
\section{Squark mixing with flavour violation}
\label{sec:sq.matrix}

In the MSSM the most general form of the squark mass matrices in the super-CKM basis of $\sq_{0 \gamma} =
(\sq_{1 {\rm L}}, \sq_{2 {\rm L}}, \sq_{3 {\rm L}}$,
$\sq_{1 {\rm R}}, \sq_{2 {\rm R}}, \sq_{3 {\rm R}}),~\gamma = 1,...6,$  
with $(q_1, q_2, q_3)=(u, c, t),$ $(d, s, b)$ is~\cite{Allanach:2008qq}
\begin{equation}
    {\cal M}^2_{\tilde{q}} = \left( \begin{array}{cc}
        {\cal M}^2_{\tilde{q},LL} & {\cal M}^2_{\tilde{q},LR} \\[2mm]
        {\cal M}^2_{\tilde{q},RL} & {\cal M}^2_{\tilde{q},RR} \end{array} \right),
 \label{EqMassMatrix}
\end{equation}
for $\tilde{q}=\tilde{u},\tilde{d}$, where the $3\times3$ matrices read
\begin{eqnarray}
    & &{\cal M}^2_{\tilde{u},LL} = V_{\rm CKM} M_Q^2 V_{\rm CKM}^{\dag} + D_{\tilde{u},LL}{\bf 1} + \hat{m}^2_u, \nonumber \\
    & &{\cal M}^2_{\tilde{u},RR} = M_U^2 + D_{\tilde{u},RR}{\bf 1} + \hat{m}^2_u, \nonumber \\
    & & {\cal M}^2_{\tilde{d},LL} = M_Q^2 + D_{\tilde{d},LL}{\bf 1} + \hat{m}^2_d,  \nonumber \\
    & & {\cal M}^2_{\tilde{d},RR} = M_D^2 + D_{\tilde{d},RR}{\bf 1} + \hat{m}^2_d.
     \label{EqM2LLRR}
\end{eqnarray}
Here $M_{Q,U,D}$ are the hermitian soft SUSY-breaking mass matrices of the squarks and
$\hat{m}_{u,d}$ are the diagonal mass matrices of the up-type and down-type quarks.
$D_{\tilde{q},LL} = \cos 2\beta m_Z^2 (T_3^q-e_q
\sin^2\theta_W)$ and $D_{\tilde{q},RR} = e_q \sin^2\theta_W \times$ $ \cos 2\beta m_Z^2$,
where
$T_3^q$ and $e_q$ are the isospin and
electric charge of the quarks (squarks), respectively, and $\theta_W$ is the weak mixing
angle.
The left-left blocks of up-type and down-type squarks are related
by the CKM matrix $V_{\rm CKM}$ due to the $SU(2)_{\rm L}$ symmetry
The off-diagonal blocks of eq.~(\ref{EqMassMatrix}) read
\begin{eqnarray}
 {\cal M}^2_{\tilde{u},RL} = {\cal M}^{2\dag}_{\tilde{u},LR} &=&
\frac{v_2}{\sqrt{2}} T^T_U - \mu^* \hat{m}_u\cot\beta, \nonumber \\
 {\cal M}^2_{\tilde{d},RL} = {\cal M}^{2\dag}_{\tilde{d},LR} &=&
\frac{v_1}{\sqrt{2}} T^T_D - \mu^* \hat{m}_d\tan\beta,
\end{eqnarray}
where $T^T_{U,D}$ are the transposes of the soft SUSY-breaking trilinear 
coupling matrices of the up-type and down-type squarks $T_{U,D}$ defined as 
${\cal L}_{int} \supset -(T_{U\alpha \beta} \ti{u}^\dagger _{R\beta}\ti{u}_{L\alpha}H^0_2 $ 
$+ T_{D\alpha \beta} \ti{d}^\dagger _{R\beta}\ti{d}_{L\alpha}H^0_1)$,
$\mu$ is the higgsino mass parameter, and $\tan\beta=v_2/v_1$, where $v_{1,2}=\sqrt{2} \left\langle H^0_{1,2} \right\rangle$
are the vacuum expectation values of the neutral Higgs fields.
The squark mass matrices are diagonalized by the $6\times6$ unitary matrices $R^{\tilde{q}}$,
$\tilde{q}=\tilde{u},\tilde{d}$, such that
\begin{eqnarray}
&&R^{\tilde{q}} {\cal M}^2_{\tilde{q}} (R^{\tilde{q} })^{\dag} = {\rm diag}(m_{\tilde{q}_1}^2,\dots,m_{\tilde{q}_6}^2) 
\end{eqnarray}
with $m_{\tilde{q}_1} < \dots < m_{\tilde{q}_6}$.
The physical mass eigenstates $\sq_i, i=1,...,6$ are given by $\sq_i =  R^{\sq}_{i \alpha} \sq_{0\alpha} $.

We define the QFV parameters in the up-type squark sector 
$\delta^{LL}_{\alpha\beta}$, $\delta^{uRR}_{\alpha\beta}$
and $\delta^{uRL}_{\alpha\beta}$ $(\alpha \neq \beta)$ as follows \cite{Gabbiani:1996hi}:
\begin{eqnarray}
\delta^{LL}_{\alpha\beta} & \equiv & M^2_{Q \alpha\beta} / \sqrt{M^2_{Q \alpha\alpha} M^2_{Q \beta\beta}}~,
\label{eq:InsLL}\\[3mm]
\delta^{uRR}_{\alpha\beta} &\equiv& M^2_{U \alpha\beta} / \sqrt{M^2_{U \alpha\alpha} M^2_{U \beta\beta}}~,
\label{eq:InsRR}\\[3mm]
\delta^{uRL}_{\alpha\beta} &\equiv& (v_2/\sqrt{2} ) T_{U\beta\alpha} / \sqrt{M^2_{U \alpha\alpha} M^2_{Q \beta\beta}}~.
\label{eq:InsRL}
\end{eqnarray}
Here $\alpha,\beta=1,2,3 ~(\alpha \ne \beta)$ denote the quark flavours $u,c,t$.
The QFV parameters relevant for this study are $\delta^{uRL}_{23}$, 
$\delta^{uLR}_{23} \equiv ( \delta^{uRL}_{32})^*$, $\delta^{uRR}_{23}$, 
and $\delta^{LL}_{23}$, which are the $\ti{c}_R - \ti{t}_L$, 
$\ti{c}_L - \ti{t}_R$, $\ti{c}_R-\ti{t}_R$, and $\ti{c}_L - \ti{t}_L$ 
mixing parameters, respectively.
We also use the QFC parameter $\delta^{uRL}_{33}$ which is defined 
by eq.~(\ref{eq:InsRL}) with $\alpha= \beta = 3$ and is the 
$\ti{t}_L - \ti{t}_R$ mixing parameter. 
We assume all QFV parameters and $\delta^{uRL}_{33} $ to be real.

 %
%
\section{QFV bosonic decays of up-type squarks}
\label{sec:bos_sq_decays}

If kinematically allowed, the following QFV bosonic decays of up-type squarks are possible:
\begin{eqnarray}
&& \su_i \to \su_j + h^0, H^0, A^0 
\label{bdec1}\\[2mm]
&&  \su_i \to \sd_j + H^+
\label{bdec2}\\[2mm]
&& \su_i \to \su_j + Z^0
\label{bdec3}\\[2mm]
&& \su_i \to \sd_j + W^+
\label{bdec4}
\end{eqnarray}
with $i, j$ = 1,...,6 specifying the squark mass eigenstates which are mixtures of the 
squark flavour eigenstates (see Section~\ref{sec:sq.matrix}). Here $h^0 (H^0)$ is the lighter (heavier) CP-even neutral Higgs boson, $A^0$ is 
the CP-odd neutral Higgs boson, and $H^+$ is the charged Higgs boson. Of course, there are also QFC bosonic squark decays. 
In this article we study mainly $\su_{2}$ 
decays in scenarios where their decays into charged bosons of eqs.\ (\ref{bdec2}) and 
(\ref{bdec4}) and those into the heavier Higgs bosons $H^0$ and $A^0$ are kinematically forbidden. The couplings between $\su_i-\su_j/\sd_j$ and 
the bosons in eqs.~(\ref{bdec1}) -- (\ref{bdec4}), taking into account QFV, are given in 
\cite{Bruhnke:2010rh}. For completeness, the couplings to the lightest Higgs 
boson, $h^0$, are listed in Appendix~\ref{sec:sq.coupl}. Note that the QFV parts are 
proportional to the soft-SUSY-breaking trilinear coupling parameter
$T_U$. In the following discussion of the 
decays we adopt the QFV parameters $\dll, \durr, \durl, \dulr$ as defined in Section~\ref{sec:sq.matrix}. 
The parameters $\dulr, \durl$ are proportional to $T_{U_{23}}$ and $T_{U_{32}}$, respectively.
In case $\su_{1,2}$ are strong mixtures of $\sca_R-\sto_R-\sto_L$, a measurement of the 
branching ratio of the decay $\su_2 \to \su_1 h^0$ gives important information 
on the QFV trilinear coupling $T_{U32}$ (i.e. $ \sca_R^{\dagger}-\sto_L-H_2^0$ coupling).

In the calculation of the branching ratios of the decays (\ref{bdec1}) -- (\ref{bdec4}) 
we have to take into account both QFV and QFC fermionic squark decays \cite{Bartl:2010du,Bartl:2009au}
\begin{eqnarray}
&& \su_i \to u_{\alpha} + \nt_k
\label{fdec1}\\[2mm]
&&  \su_i \to d_{\alpha} + \chp_l
\label{fdec2}\\[2mm]
&&  \su_i \to u_{\alpha} + \sg
\label{fdec3}
\end{eqnarray}
where $\alpha=1,2,3$ is the flavour index, $\nt_k, k=1,...,4$, are the neutralinos and $\chp_l, l=1,2$, are the charginos. 
As $\su_{1,2}$ are mainly mixtures of $\sca_R$, $\sto_R$ and $\sto_L$ in the scenarios under consideration,       
both decays $\su_{1,2} \to t \nt_1$ and $\su_{1,2} \to c \nt_1$ are possible.  
%
\begin{table}[h!]
\caption{Weak scale basic MSSM parameters at $Q=1~{\rm TeV}$~\cite{AguilarSaavedra:2005pw} for scenario A, 
except for $m_{A^0}$ which is the pole mass (i.e.\ the physical mass) 
of $A^0$. All of $T_{U \alpha \alpha}$ and $T_{D \alpha \alpha}$ are zero, except for 
$T_{U33} = - 2160$~GeV (i.e. $\delta^{uRL}_{33} = - 0.34$). All other squark 
parameters not shown here are zero. }
\begin{center}
\begin{tabular}{|c|c|c|}
  \hline
 $M_1$ & $M_2$ & $M_3$ \\
 \hline \hline
 400~\gev  &  800~\gev &  1000~\gev \\
  \hline
\end{tabular}
\vskip0.4cm
\begin{tabular}{|c|c|c|}
  \hline
 $\mu$ & $\tan \beta$ & $m_{A^0}$ \\
 \hline \hline
 2640~\gev & 20 &  1500~\gev \\
  \hline
\end{tabular}
\vskip0.4cm
\begin{tabular}{|c|c|c|c|}
  \hline
   & $\alpha = 1$ & $\alpha= 2$ & $\alpha = 3$ \\
  \hline \hline
   $M_{Q \alpha \alpha}^2$ & $(2400)^2~\gev^2$ &  $(2360)^2~\gev^2$  & $(1450)^2~\gev^2$ \\
   \hline
   $M_{U \alpha \alpha}^2$ & $(2380)^2~\gev^2$ & $(780)^2~\gev^2$ & $(750)^2~\gev^2$ \\
   \hline
   $M_{D \alpha \alpha}^2$ & $(2380)^2~\gev^2$ & $(2340)^2~\gev^2$ &  $(2300)^2~\gev^2$  \\
   \hline
\end{tabular}
\vskip0.4cm
\begin{tabular}{|c|c|c|c|}
  \hline
   $\delta^{LL}_{23}$ & $\delta^{uRR}_{23}$  &  $\delta^{uRL}_{23}$ & $\delta^{uLR}_{23}$\\
  \hline \hline
   0 & 0.3 &  -0.07   &  0  \\
    \hline
\end{tabular}
\end{center}
\label{basicparam}
\end{table}
\begin{table}[h!]
\caption{Physical masses in GeV of the particles in scenario A (see Table~\ref{basicparam}).}
\begin{center}
\begin{tabular}{|c|c|c|c|c|c|}
  \hline
  $\mnt{1}$ & $\mnt{2}$ & $\mnt{3}$ & $\mnt{4}$ & $\mch{1}$ & $\mch{2}$ \\
  \hline \hline
  $397$ & $824$ & $2623$ & $2625$ & $825$ & $2625$ \\
  \hline
\end{tabular}
\vskip 0.4cm
\begin{tabular}{|c|c|c|c|c|}
  \hline
  $m_{h^0}$ & $m_{H^0}$ & $m_{A^0}$ & $m_{H^+}$ \\
  \hline \hline
  $124.0$  & $1496$ & $1500$ & $ 1510$ \\
  \hline
\end{tabular}
\vskip 0.4cm
\begin{tabular}{|c|c|c|c|c|c|c|}
  \hline
  $\msg$ & $\msu{1}$ & $\msu{2}$ & $\msu{3}$ & $\msu{4}$ & $\msu{5}$ & $\msu{6}$ \\
  \hline \hline
  $1141$ & $605$ & $861$ & $1477$ & $2387$ & $2401$ & $2427$ \\
  \hline
\end{tabular}
\vskip 0.4cm
\begin{tabular}{|c|c|c|c|c|c|}
  \hline
  $\msd{1}$ & $\msd{2}$ & $\msd{3}$ & $\msd{4}$ & $\msd{5}$ & $\msd{6}$ \\
  \hline \hline
 $1433$ & $2321$ & $2364$ & $2388$ & $2404$ & $2428$ \\
  \hline
\end{tabular}

\end{center}
\label{physmasses}
\end{table}
%
%
\begin{table}[h!]
\caption{Flavour decomposition of $\su_1$ and $\su_2$ in scenario A of Table~\ref{basicparam}. Shown are the squared coefficients. }
\begin{center}
\begin{tabular}{|c|c|c|c|c|c|c|c|}
  \hline
  & $\su_L$ & $\sca_L$ & $\sto_L$ & $\su_R$ & $\sca_R$ & $\sto_R$ \\
  \hline \hline
 $\su_1$  & $0$ & $0$ & $0.032$ & $0$ & $0.209$ & $0.759$ \\
  \hline
  $\su_2$  & $0$ & $0$ & $0.031$ & $0$ & $0.785$ & $0.184$ \\
  \hline
\end{tabular}
\end{center}
\label{flavourdecomp}
\end{table}
\begin{table}[h!]
\caption{Two-body decay branching ratios of $\su_2$, $\su_1$ and gluino in scenario A of Table~\ref{basicparam}. The charge conjugated  
processes have the same branching ratios and are not shown explicitly.}
\begin{center}
\begin{tabular}{|c|c|}
  \hline
   B$(\su_2 \to  \su_1 h^0)$ & 0.47\\
   \hline
    B$(\su_2 \to \su_1 Z^0)$ & 0.01\\
   \hline
   B$(\su_2 \to c \nt_1)$ & 0.43\\
   \hline
   B$(\su_2 \to t \nt_1)$ & 0.09\\
   \hline \hline
  B$(\su_1 \to c \nt_1)$ & 0.36\\
  \hline
  B$(\su_1 \to t \nt_1)$ & 0.64\\
   \hline \hline
   B$(\sg \to  \su_2 \bar{c})$ & 0.12\\
   \hline
   B$(\sg \to  \su_2 \bar{t})$ & 0.01\\
   \hline
     B$(\sg \to  \su_1 \bar{c})$ & 0.09\\
   \hline
   B$(\sg \to  \su_1 \bar{t})$ & 0.27\\
   \hline
   \end{tabular}
\end{center}
\label{BRsA}
\end{table}
\begin{figure*}
\centering
\subfigure[]{
   { \mbox{\hspace*{-1cm} \resizebox{8.cm}{!}{\includegraphics{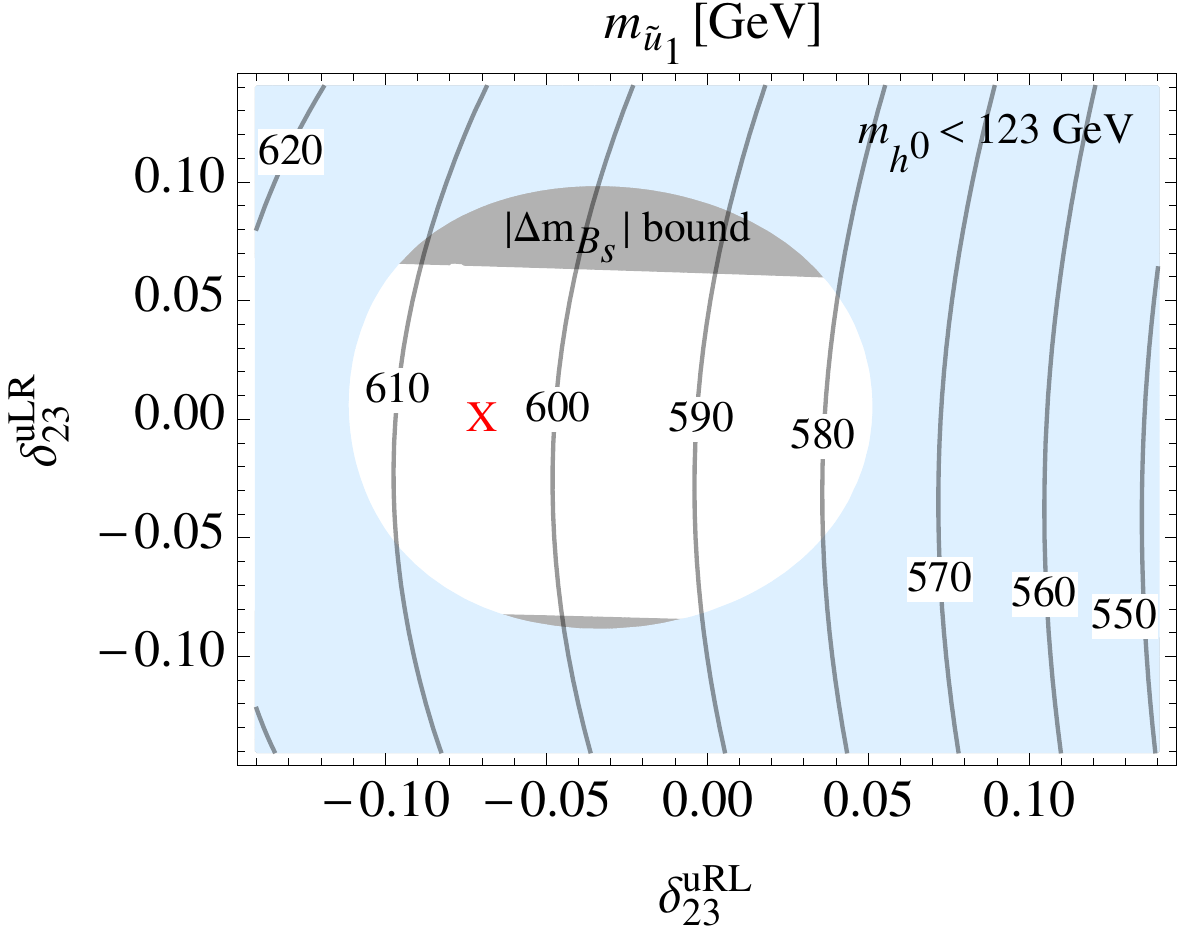}} \hspace*{-0.8cm}}}
   \label{msu1}}
 \subfigure[]{
   { \mbox{\hspace*{+0.cm} \resizebox{8.cm}{!}{\includegraphics{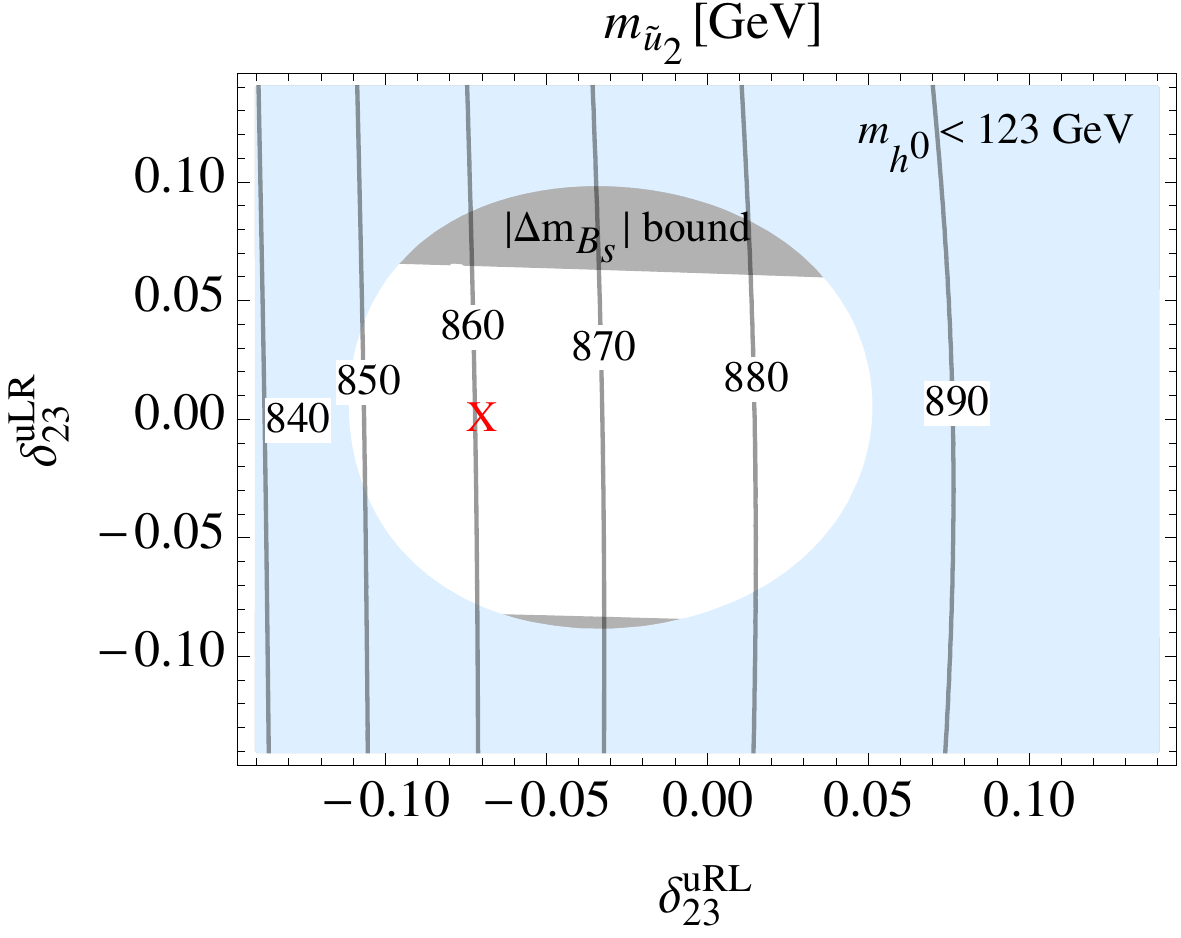}} \hspace*{-1cm}}}
  \label{msu2}}
\caption{Dependence of the masses of $\su_1$ (a) and $\su_2$ (b) on $\durl$ and $\dulr$ where the other parameters are fixed as in Table~\ref{basicparam} and "X" in both plots corresponds
to scenario A.
\label{msu12}}
\end{figure*}
\begin{figure*}[h!]
\centering
\subfigure[]{
   { \mbox{\hspace*{-1cm} \resizebox{8.cm}{!}{\includegraphics{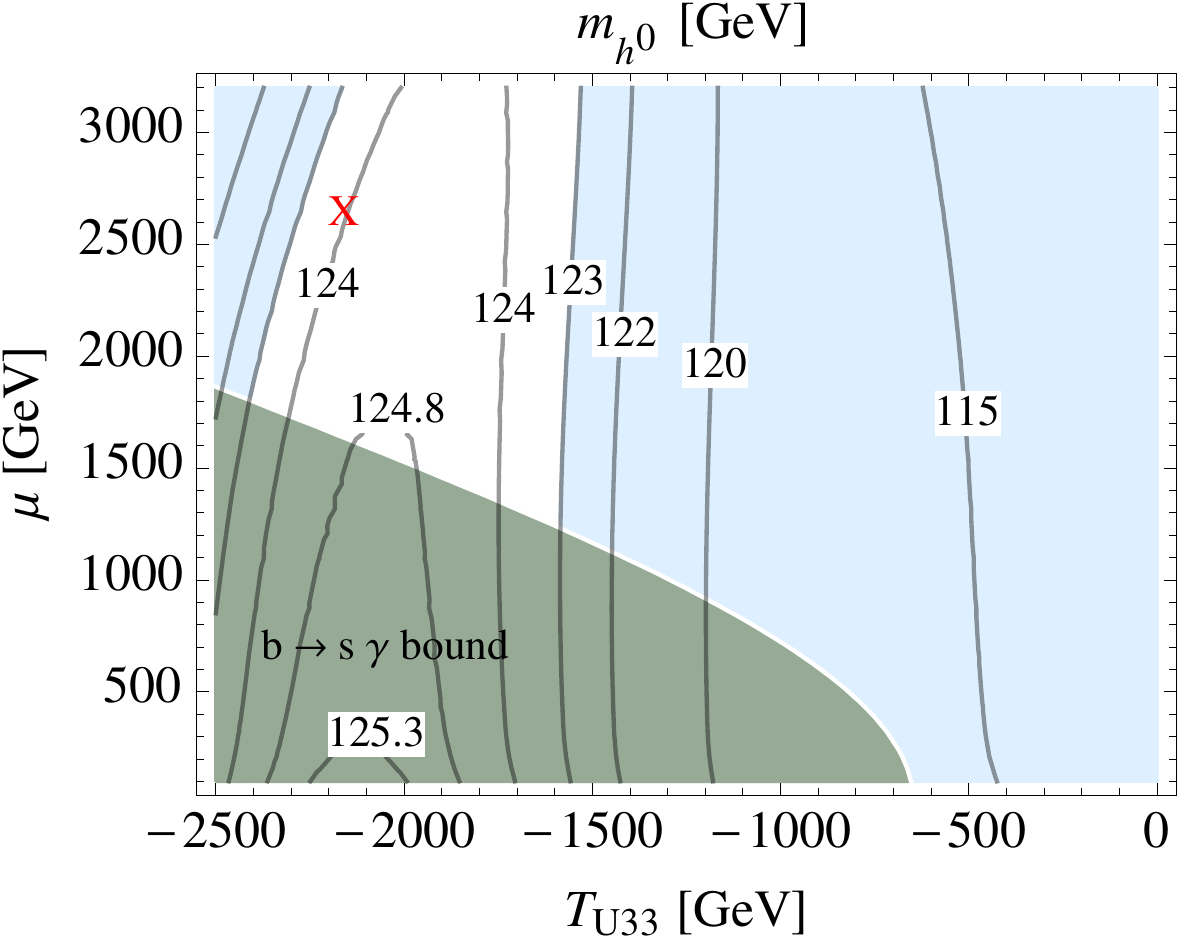}} \hspace*{-0.8cm}}}
   \label{mh0a}}
 \subfigure[]{
   { \mbox{\hspace*{+0.cm} \resizebox{8.cm}{!}{\includegraphics{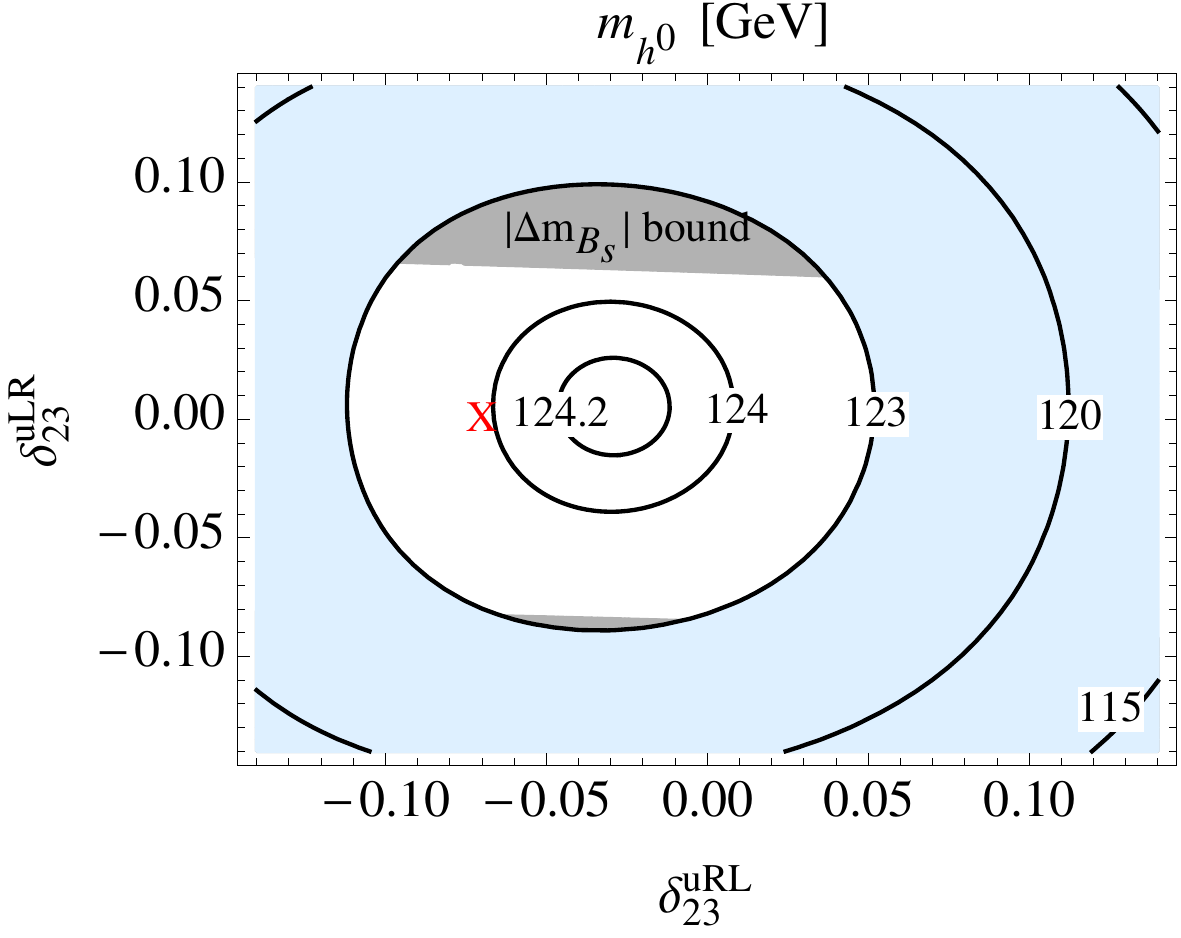}} \hspace*{-1cm}}}
  \label{mh0b}}
\caption{The mass of the lightest Higgs boson, $m_{h^0}$, as a function of $T_{U33}$ and $\mu$ (a)  and as a function of $\durl$ and $\dulr$ (b) where the other parameters are fixed as in Table~\ref{basicparam} and "X" in both plots corresponds
to scenario A. The light shaded (light blue) areas indicate $m_{h^0}<123~\gev$ (see Table~\ref{TabConstraints}).
 \label{mh0}}
\end{figure*}

In the following we first study the QFV bosonic decays in detail
    for the scenario with the parameters in Table~\ref{basicparam} (scenario A).
    In a second step we consider two variants of this scenario which contain
    substantial new features. The parameters are chosen such that the two
    lightest $u$-squarks and the gluino can be produced with sizable rates
    at the LHC with $\sqrt{s}=14$ TeV.
Moreover, the mass difference between $\su_2$ and $\su_1$ is such that the decay $\su_2 \to \su_1 h^0$ is kinematically possible. We choose relatively large values of $M_1, M_2$ and $\mu$ in order to avoid the dominance of the fermionic squark decays of $\su_2$. The hierarchy between the values of $M^2_{Q\alpha\alpha}$ and $M^2_{U\alpha\alpha}, \alpha=1,2,3$, is chosen to allow sizable $\tilde{c}_R - \tilde{t}_L$ mixing effects.
For this scenario all experimental and theoretical constraints given in Appendix~\ref{sec:exp_constr} are satisfied.
In particular, $\msg, \msq{}$ (for the first and the second generation) and $\mnt{1}$ obey the experimental bounds. For the low-energy observables we obtain the following values:
$\Delta M_{B_s}=16.4$ ps$^{-1}$, 
B$(b \to s \gamma)=3.0 \cdot 10^{-4}$, B($B_s \to \mu^+ \mu^-)=3.3 \cdot 10^{-9}$,
B($B_u \to \tau \nu_\tau)= 1.08 \cdot 10^{-4}$.
All numerical calculations in this study, except for the cross
sections, are performed with the 
public code SPheno
v3.2~\cite{Porod:2003um,
Porod:2011nf}. 
In the calculation of the low energy observables large chirally
enhanced corrections may be important see e.g.~\cite{Carena:2000uj,Buras:2002vd,Crivellin:2011jt}. Using the program SUSY\_FLAVOR v2.10~\cite{Crivellin:2012jv} we have calculated the low-energy B observables in our scenarios and compared them with the results obtained with SPheno. We have found agreement within 10\%. The resummation effect of the chirally enhanced corrections in SUSY\_FLAVOR v2.10 is less than 
1\% in the scenarios considered.
We also use the package SSP~\cite{Staub:2011dp} that allows an efficient handling of parameter studies. 
The physical masses of squarks, gluino, charginos, neutralinos and Higgs bosons are shown in Table~\ref{physmasses}. 
We obtain 
$m_{h^0}=124$ GeV which is in the range of the Higgs signal at LHC~\cite{CMS@ICHEP2012,ATLAS@ICHEP2012, ATLAS:2012ae, Aad:2012tfa,Chatrchyan:2012tx,Chatrchyan:2012ufa}.
 Moreover, in this scenario we are in the decoupling limit with $m_{A^0}=1500$ GeV $\gg m_{h^0}$, and hence the lightest Higgs boson $h^0$ is SM-like.
The flavour decompositions of $\su_1$ and $\su_2$ in scenario A are shown in Table~\ref{flavourdecomp}. 
In this scenario with large $\delta^{uRR}_{23}$, $\delta^{uRL}_{33}$ and 
$\delta^{uRL}_{23}$, $\su_1$ is mainly a $\sto_R-\sca_R (-\sto_L)$ 
mixture and $\su_2$ is mainly a $\sca_R-\sto_R (-\sto_L)$ mixture.

We studied the QFV fermionic decays of gluinos and squarks in~\cite{Bartl:2010du, Bartl:2009au, Bartl:2011wq}. There it turned out that QFV effects mainly depend on $\durr$ and $\ddrr$, whereas the influence of the other QFV parameters is much weaker. 
In Table~\ref{basicparam} we have taken $\durr=0.3$ in order to have the branching ratios for the QFV decays $\su_1 \to c \nt_1$ and $\su_1 \to t \nt_1$ comparable.
In the present paper we concentrate on the dependence of the QFV effects on $\durl$ and $\dulr$, which enter the squark-squark-Higgs couplings.

In Fig.~\ref{msu1} and~\ref{msu2} we show the mass contours of $\su_1$ and $\su_2$ in the $\durl -\dulr$ plane for scenario A with all other parameters as in Table~\ref{basicparam}.
In all contour plots in this article the white regions satisfy all the
experimental and theoretical constraints listed in~\ref{sec:exp_constr}.
One can see a somewhat stronger dependence on $\durl$ due to 
the sizable mass-splitting induced by the $\sca_R-\sto_L$ mixing, 
which is a consequence of the chosen hierarchy within $M^2_Q$ 
and $M^2_U$.

In Fig.~\ref{mh0a} the lightest Higgs mass $m_{h^0}$ is shown 
as a function of $T_{U33}$ and $\mu$. In order to obtain  
$m_{h^0}$ within the allowed parameter range a large $|T_{U33}|$ 
and a rather large $\mu$~\cite{Brummer:2012ns, Arbey:2012dq, Benbrik:2012rm} are required.
A large $|T_{U33}|$ enhances the stop loop corrections and a large $|\mu|$ enlarges the sbottom loop corrections through the 
term $A_b-\mu \tan \beta$.
In Fig.~\ref{mh0b} we show  
contours of $m_{h^0}$ in the $\dulr$ - $\durl$ plane. Within the shown range of the QFV 
parameters, $m_{h^0}$ varies 
by about 1.5 GeV due to the $\sca$ admixture in the stop loops.
%
\begin{figure*}
\centering
   { \mbox{\hspace*{-1cm} \resizebox{8.3cm}{!}{\includegraphics{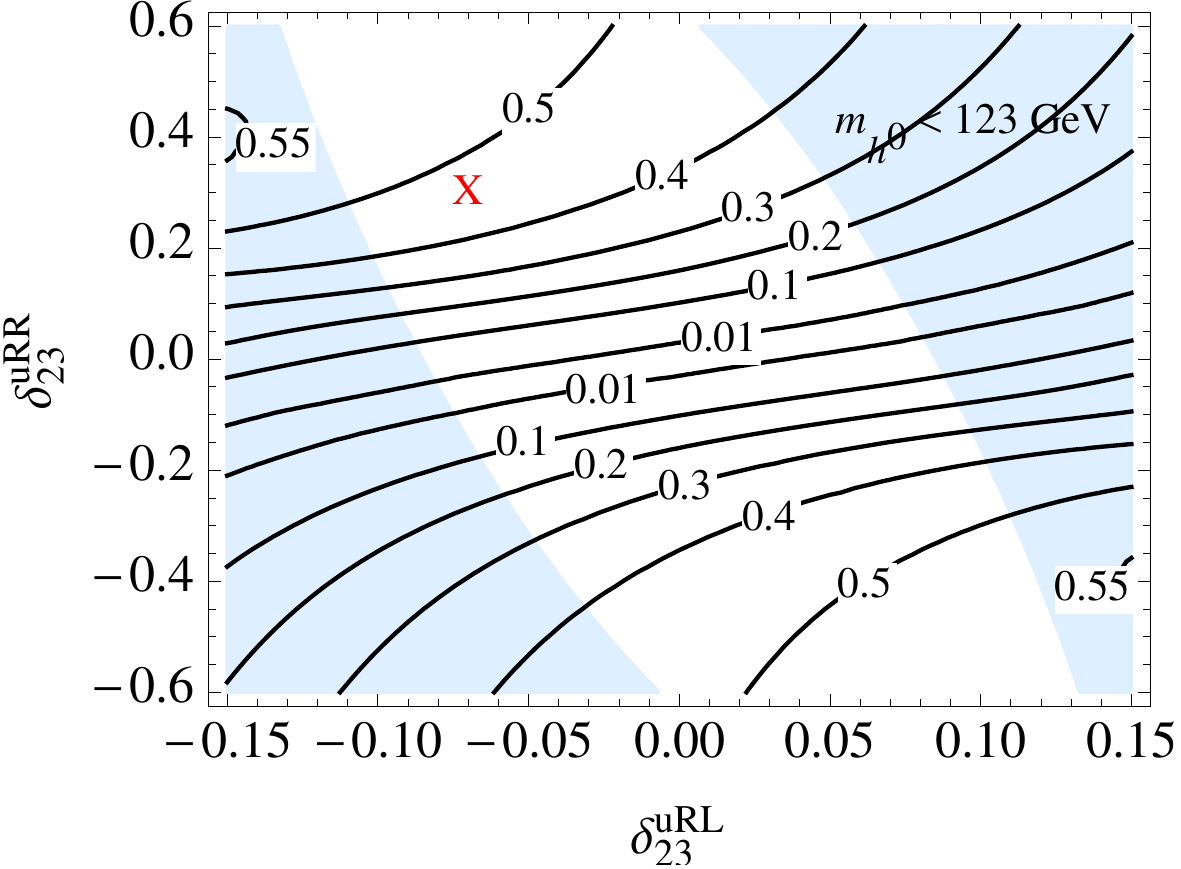}}\hspace*{-0.8cm}}}
\caption{The branching ratio B$(\su_2 \to \su_1 h^0)$ as a function of $\durl$  and $\durr$ in scenario A. 
"X" indicates the reference point defined with 
the parameters of Table~\ref{basicparam}.}
\label{BRsu2tosu1h0scA}
\end{figure*}

\begin{figure*}
\centering
\subfigure[]{
   { \mbox{\hspace*{-1cm} \resizebox{8.cm}{!}{\includegraphics{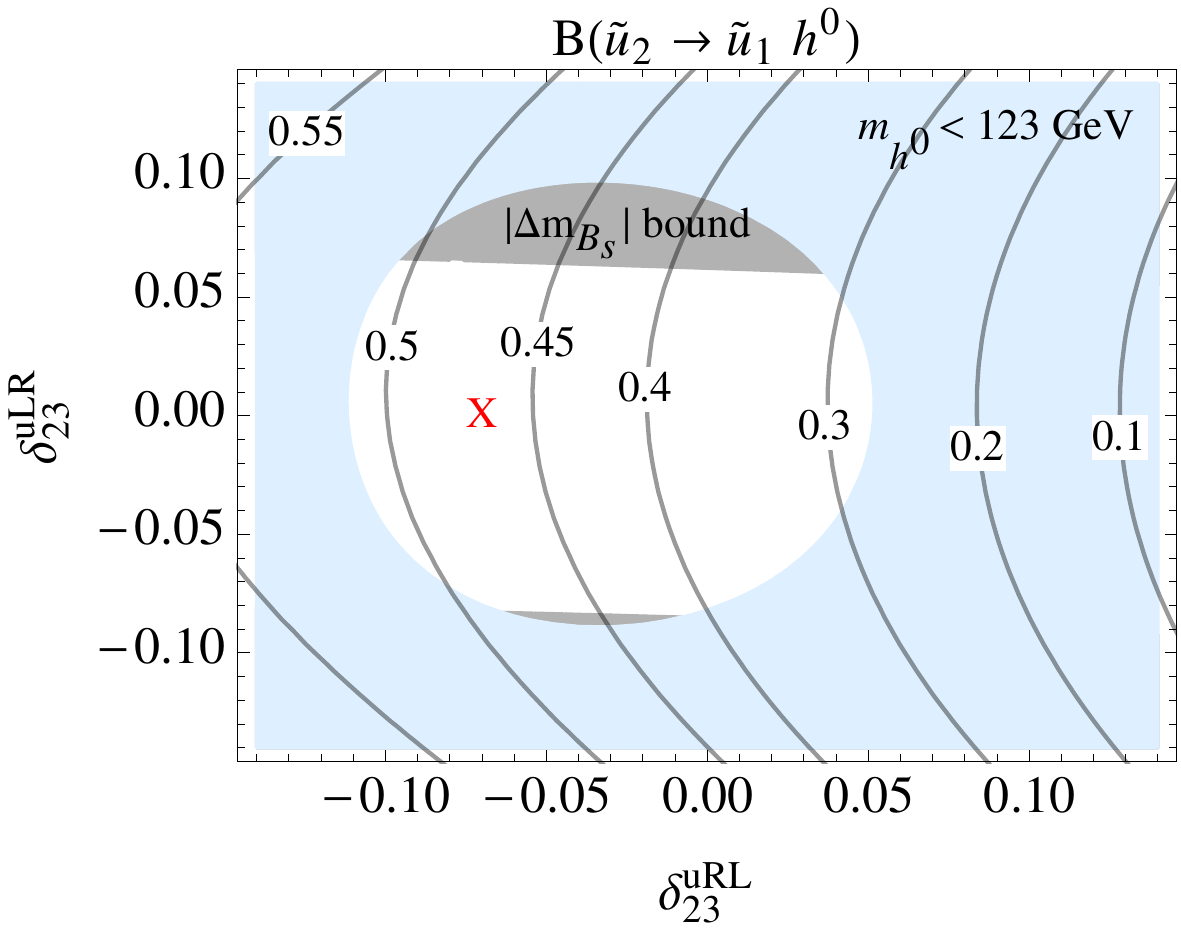}} \hspace*{-0.8cm}}}
  \label{BRsu2tosu1h0a}}
 \subfigure[]{
   { \mbox{\hspace*{+0.cm} \resizebox{8.cm}{!}{\includegraphics{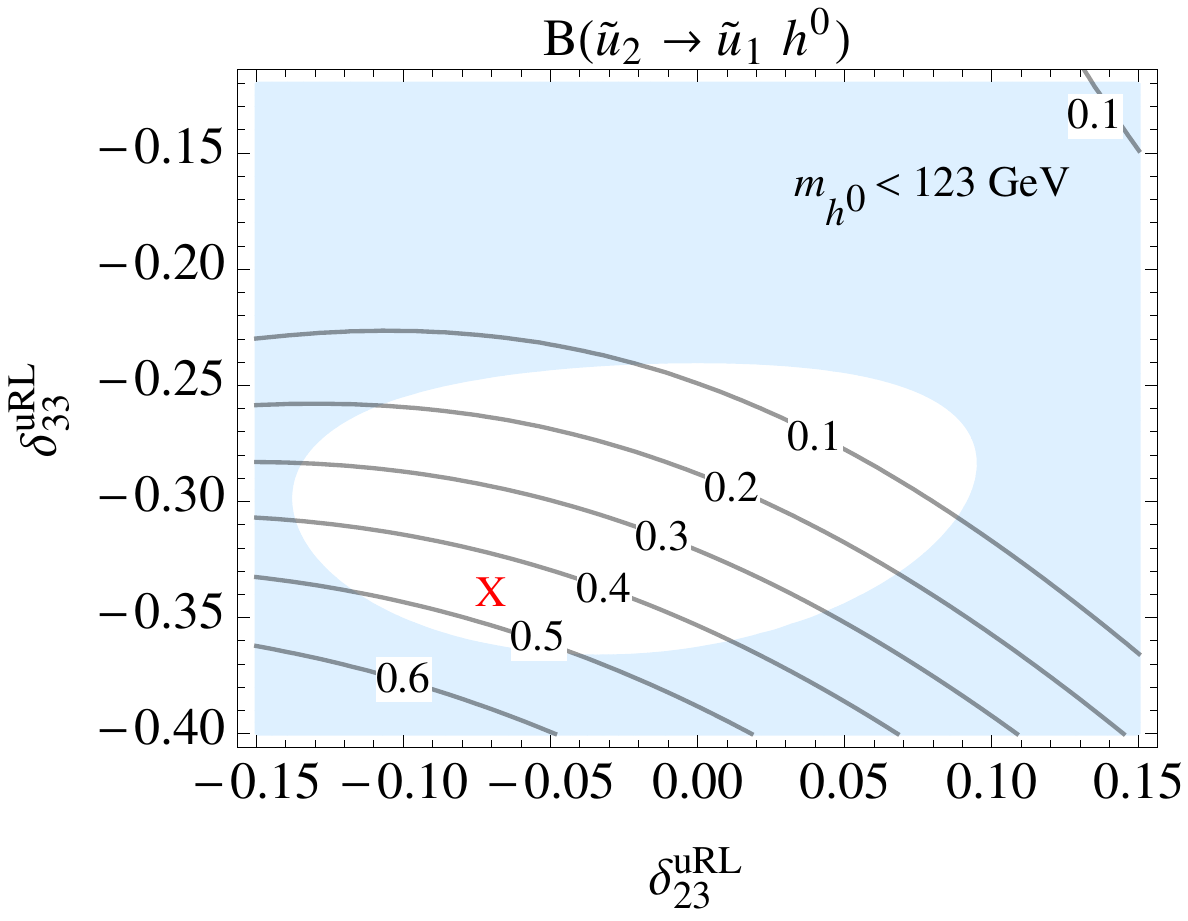}} \hspace*{-1cm}}}
  \label{BRsu2tosu1h0b}}
\caption{The branching ratio of the decay $\su_2 \to \su_1 h^0$ as a function of $\durl$ and $\dulr$ (a) and as a function of $\durl$ and $\delta^{uRL}_{33}$ (b) with the other parameters fixed as in Table~\ref{basicparam} and "X" in both plots corresponds
to scenario A.
 \label{BRsu2tosu1h0}}
\end{figure*}
%

Next we study the decay $\su_2 \to \su_1 h^0$ in more detail. 
In Fig.~\ref{BRsu2tosu1h0scA} we show the branching ratio B$(\su_2 \to \su_1 h^0)$ 
as a function of $\durl$ and $\durr$. 
Note that $\durr$ must be different from 0 in order to have this 
branching ratio sizable.
At the reference point of scenario A (see Table~\ref{basicparam}) 
B$(\su_2 \to \su_1 h^0) \approx 0.45$. The $\durl$ and $\delta^{uRR}_{23}$
dependences can be understood by the arguments below. Note that 
$\su_{1,2}$ become strong mixtures of $\sca_R$ and $\sto_R$ for sizable $\delta^{uRR}_{23}$.  

\begin{figure*}
\centering
\subfigure[]{
   {\mbox{\hspace*{-1cm} \resizebox{8.cm}{!}{\includegraphics{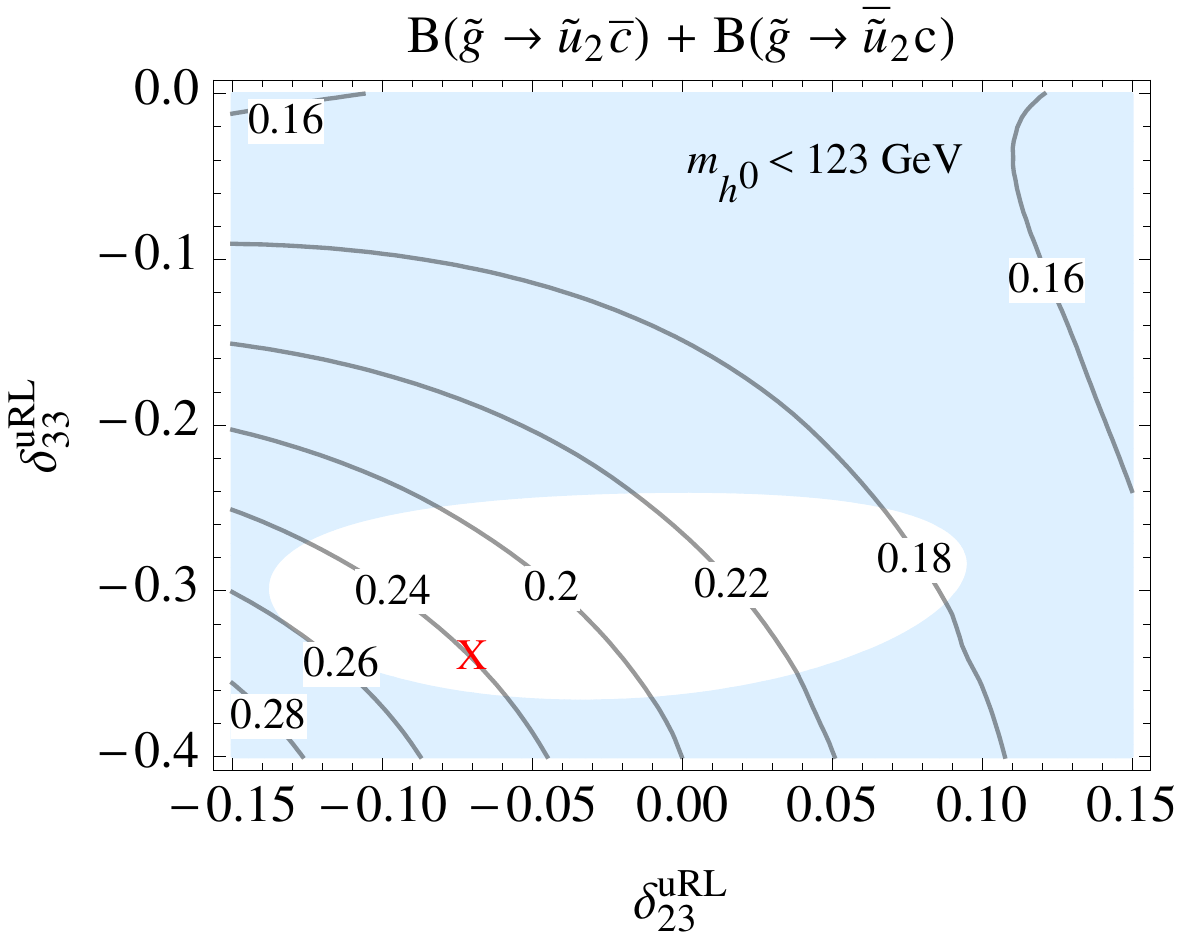}} \hspace*{-0.8cm}}}
   \label{BRsgtosu2c}}
 \subfigure[]{
   { \mbox{\hspace*{+0.cm} \resizebox{8.cm}{!}{\includegraphics{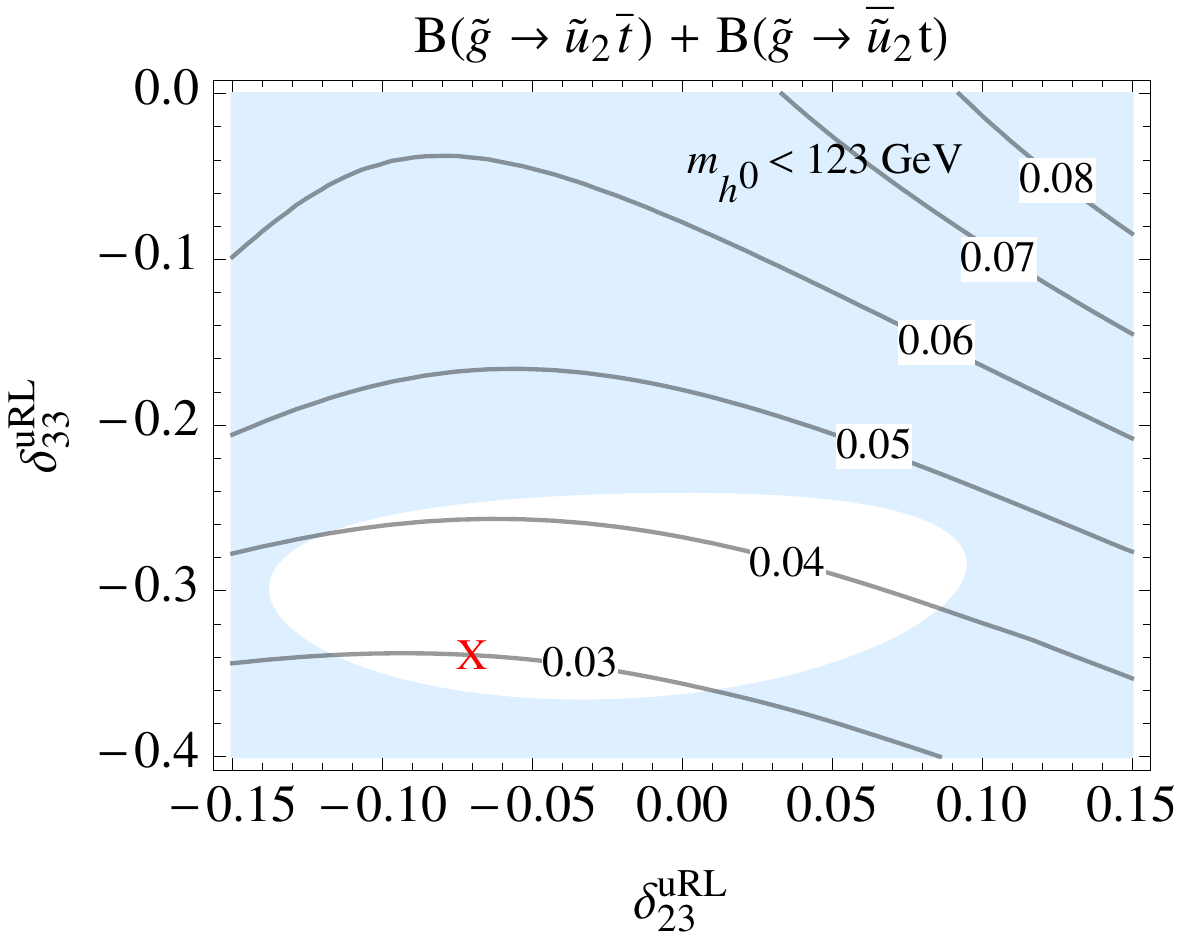}} \hspace*{-1cm}}}
  \label{BRsgtosu2t}}
\caption{The branching ratios B$(\sg \to \su_2 \bar{c})+$B$(\sg \to \bar{\su}_2 c)$ (a) and B$(\sg \to \su_2 \bar{t})+$B$(\sg \to \bar{\su}_2 t)$ (b) as functions of $\durl$ and $\delta^{uRL}_{33}$ with the other parameters fixed as in Table~\ref{basicparam} and "X" in both plots corresponds
to scenario A.
\label{BRsgtosu2ct}}
\end{figure*}

In Fig.~\ref{BRsu2tosu1h0a} we show the branching ratio of the decay $\su_2 \to \su_1 h^0$ as a function of the 
QFV parameters $\durl$ and $\dulr$. 
We have used the formulas eqs.~(\ref{eqSuSuh0}), (\ref{coupluiujh}) and (\ref{decratesuisujh0}) of Appendix~\ref{sec:sq.coupl}. The dominant terms in the 
coupling $c_{\su_2 \su_1 h^0}$ are those proportional to $T_{U33}$ and $T_{U32}$, since $\su_{1,2}$ are mainly mixtures of 
$\sca_R, \sto_R$ and $\sto_L$ (recall that $\durl \sim T_{U32}, ~\dulr \sim T_{U23}, ~\delta^{uRL}_{33} \sim T_{U33}$, see eq.~(\ref{eq:InsRL}) ). 
The term proportional to $ T_{U23}$ is rather small due to the small $\sca_L$ components of $\su_{1,2}$. 
Therefore, the dependence on $\durl$  is 
stronger than that on $\dulr$. In the region allowed by the constraints listed in Appendix~\ref{sec:exp_constr} the 
branching ratio for $\su_2 \to \su_1 h^0$ can go up to 50\%. 
The decrease of B($\su_2 \to \su_1 h^0$) with increasing $\durl$ is 
due to an interference of the dominant terms with the couplings $T_{U33}$
and $T_{U32}$ in eq.~(\ref{coupluiujh}).

In Fig.~\ref{BRsu2tosu1h0b} we show the branching ratio B($\su_2 \to \su_1 h^0$) as a function of the QFV parameter $\durl$ and the QFC parameter $\delta^{uRL}_{33}$ 
(recall that $\delta^{uRL}_{33}$ corresponds to the $\sto_L-\sto_R$ mixing). 
The decrease of B($\su_2 \to \su_1 h^0$) with increasing $\durl$ is due to the same reason as in Fig.~\ref{BRsu2tosu1h0a}. 
B($\su_2 \to \su_1 h^0$) grows with increasing $|\delta^{uRL}_{33}|$ 
because $T_{U33}$ as well as the $\sto_L$ components of $\su_{1,2}$ 
become larger. 

In Table~\ref{BRsA} we give the branching ratios for the two-body decays of $\su_2, \su_1$ and gluino at the reference point of scenario A. 
Note that the $\sca_R-\sto_R$ mixing and the $\sca_R-\sto_L$ mixing together with the large top trilinear coupling $T_{U33}$ 
lead to an enhanced branching ratio for $\su_2 \to \su_1 h^0$.

We have also studied the QFV decays $\su_3 \to \su_1 h^0$ and $\su_3 \to \su_2 h^0$. The branching ratio of $\su_3 \to \su_1 h^0$ can go up to $30\%$. At the reference point of Table~\ref{basicparam} it is about $23\%$. $\su_3$ has a large $\sto_L$ component. Hence, the decay $\su_3 \to \su_1 h^0$ is mainly due to $\sto_L \to \sto_R h^0$  transitions. On the other hand, in the decay  $\su_3 \to \su_2 h^0$ the behaviour of the branching ratio is very different, because the $\sto_L - \sca_R$ transitions are more important. Its branching ratio at the reference point is about $10\%$. The branching ratios of  $\su_3 \to \su_1 Z^0$ and  $\su_3 \to \su_2 Z^0$ at the reference point are about $28\%$ and $13\%$, respectively.  

In scenario A the squark $\su_2$ can also be produced in the decay of the gluino. We show in Fig.~\ref{BRsgtosu2c} and ~\ref{BRsgtosu2t} the branching ratios of the decays 
$\sg \to  \su_2 \bar{c}+c.c.$ and $\sg \to  \su_2 \bar{t}+c.c.$ as functions of $\durl$ and $\delta^{uRL}_{33}$. The branching ratio of the decay  $\sg \to  \su_2 \bar{t}+c.c.$ is much smaller than that of  $\sg \to  \su_2 \bar{c}+c.c.$ due to phase space and because the $\tilde{c}_R$ component of $\su_2$ is larger than the $\tilde{t}_{L,R}$ components.
In Table~\ref{BRsA} we give all of the branching ratios of  
gluino decays (except charge conjugate decays) for the reference 
point of scenario A.
%

\begin{table}[h!]
\caption{Two-body decay branching ratios of $\su_2$, $\su_1$ and gluino in scenario B, see Table~\ref{basicparam} and eq.~(\ref{GUTmasses}). The charge conjugated 
processes have the same branching ratios and are not shown explicitly.}
\begin{center}
\begin{tabular}{|c|c|}
  \hline
   B$(\su_2 \to  \su_1 h^0)$ & 0.39\\
   \hline
    B$(\su_2 \to \su_1 Z^0)$ & 0.01\\
   \hline
   B$(\su_2 \to c \nt_1)$ & 0.45\\
   \hline
   B$(\su_2 \to t \nt_1)$ & 0.10\\
   \hline \hline
  B$(\su_1 \to c \nt_1)$ & 0.26\\
  \hline
  B$(\su_1 \to t \nt_1)$ & 0.73\\
   \hline \hline
   B$(\sg \to  \su_2 \bar{c})$ & 0.16\\
   \hline
   B$(\sg \to  \su_2 \bar{t})$ & 0.04\\
   \hline
     B$(\sg \to  \su_1 \bar{c})$ & 0.07\\
   \hline
   B$(\sg \to  \su_1 \bar{t})$ & 0.22\\
   \hline
   \end{tabular}
\end{center}
\label{BRsB}
\end{table}
Note that in this scenario the gaugino mass parameters $M_1$,  $M_2$, and  $M_3$ do not obey the GUT relation  $M_1\approx  0.5~M_2$,  $M_3/ M_2 = g^2_3/  g^2_2$, where $g_2$ and $g_3$ are the SU(2) and SU(3) gauge coupling constants, respectively.
We define a variant of scenario A by replacing
in Table~\ref{basicparam} only the gaugino mass parameters by
\begin{equation}
M_1=250~\gev,~M_2=500~\gev,~M_3=1500~\gev
\label{GUTmasses} 
\end{equation}
which satisfy approximately the GUT relations. We call it scenario B.
The physical masses of the squarks are almost the same as in Table~\ref{physmasses}
and we do not show them explicitly.
In this scenario the gluino is relatively heavy, $m_{\sg} = 1626~\gev$, therefore, it has a 
relatively small pair production cross section $pp \to \sg \sg X$ (3.5~fb). As we will 
see in the next section, gluino production is important, because the 
lighter squarks $\su_{1,2}$ are also produced in the gluino decays 
$\sg \to \su_{1,2}~\bar{q}$. 
In this
scenario the dependences of $m_h^0$ and B$(\su_2 \to 
\su_1 h^0)$ on the QFV and QFC parameters are very similar to those of scenario A, shown in Figs.~\ref{mh0} and \ref{BRsu2tosu1h0}.
The two-body decay branching ratios of $\su_2, \su_1$ and gluino for scenario B are shown in Table~\ref{BRsB}.

\begin{table}[h!]
\caption{Physical masses in GeV of the particles in scenario C, see Table~\ref{basicparam} and eq.~(\ref{scenC}).}
\begin{center}
\begin{tabular}{|c|c|c|c|c|c|}
  \hline
  $\mnt{1}$ & $\mnt{2}$ & $\mnt{3}$ & $\mnt{4}$ & $\mch{1}$ & $\mch{2}$ \\
  \hline \hline
  $398$ & $819$ & $2623$ & $2625$ & $819$ & $2625$ \\
  \hline
\end{tabular}
\vskip 0.4cm
\begin{tabular}{|c|c|c|c|c|}
  \hline
  $m_{h^0}$ & $m_{H^0}$ & $m_{A^0}$ & $m_{H^+}$ \\
  \hline \hline
  $123.7$  & $1497$ & $1500$ & $ 1537$ \\
  \hline
\end{tabular}
\vskip 0.4cm
\begin{tabular}{|c|c|c|c|c|c|c|}
  \hline
  $\msg$ & $\msu{1}$ & $\msu{2}$ & $\msu{3}$ & $\msu{4}$ & $\msu{5}$ & $\msu{6}$ \\
  \hline \hline
  $1134$ & $651$ & $800$ & $1580$ & $2387$ & $2401$ & $2427$ \\
  \hline
\end{tabular}
\vskip 0.4cm
\begin{tabular}{|c|c|c|c|c|c|}
  \hline
  $\msd{1}$ & $\msd{2}$ & $\msd{3}$ & $\msd{4}$ & $\msd{5}$ & $\msd{6}$ \\
  \hline \hline
 $807$ & $2321$ & $2363$ & $2388$ & $2404$ & $2428$ \\
  \hline
\end{tabular}
\end{center}
\label{physmassesC}
\end{table}
%
%
\begin{table}[h!]
\caption{Flavour decomposition of $\su_1$ and $\su_2$ in scenario C, see Table~\ref{basicparam} and eq.~(\ref{scenC}). Shown are the squared coefficients. }
\begin{center}
\begin{tabular}{|c|c|c|c|c|c|c|c|}
  \hline
  & $\su_L$ & $\sca_L$ & $\sto_L$ & $\su_R$ & $\sca_R$ & $\sto_R$ \\
  \hline \hline
 $\su_1$  & $0$ & $0$ & $0.242$ & $0$ & $0.745$ & $0.012$ \\
  \hline
  $\su_2$  & $0$ & $0$ & $0.713$ & $0$ & $0.255$ & $0.032$ \\
  \hline
\end{tabular}
\end{center}
\label{flavourdecompC}
\end{table}

In scenarios A and B the decay $\su_2 \to \su_1 Z^0$ has a very small branching ratio. In the following we present a scenario (scenario C) where the branching ratios of $\su_2 \to \su_1 h^0$ and $\su_2 \to \su_1 Z^0$ are both large. For this purpose we have again changed
some of the MSSM parameters with respect to Table~\ref{basicparam}, leaving 
all other parameters unchanged,
\begin{eqnarray}
&& M_{U22}^2 = (650~\gev)^2,\quad M_{U33}^2 = (1600~\gev)^2, \quad 
M_{Q33}^2 = (780~\gev)^2, \nonumber \\
&&  \delta^{uRR}_{23} = 0, \quad \delta^{uRL}_{23} = -0.17, \quad \delta^{uRL}_{33} = - 0.3. 
\label{scenC}
\end{eqnarray}
%
\begin{table}[h!]
\caption{Two-body decay branching ratios of $\su_2$, $\su_1$ and gluino in scenario C, see Table~\ref{basicparam} and eq.~(\ref{scenC}). The charge conjugated 
processes have the same branching ratios and are not shown explicitly.}
\begin{center}
\begin{tabular}{|c|c|}
  \hline
  B$(\su_2 \to  \su_1 h^0)$ & 0.43\\
   \hline
   B$(\su_2 \to \su_1 Z^0)$ & 0.34\\
   \hline
   B$(\su_2 \to c \nt_1)$ & 0.17\\
   \hline
   B$(\su_2 \to t \nt_1)$ & 0.06\\
   \hline \hline
  B$(\su_1 \to c \nt_1)$ & 0.96\\
  \hline
   B$(\su_1 \to t \nt_1)$ & 0.04\\
   \hline \hline
   B$(\sg \to  \su_2 \bar{c})$ & 0.04\\
   \hline
   B$(\sg \to  \su_2 \bar{t})$ & 0.08\\
   \hline
   B$(\sg \to  \su_1 \bar{c})$ & 0.19\\
   \hline
   B$(\sg \to  \su_1 \bar{t})$ & 0.05\\
   \hline
   \end{tabular}
\end{center}
\label{BRsC}
\end{table}

\begin{figure}
\centering
   { \mbox{\hspace*{-1cm} \resizebox{8.3cm}{!}{\includegraphics{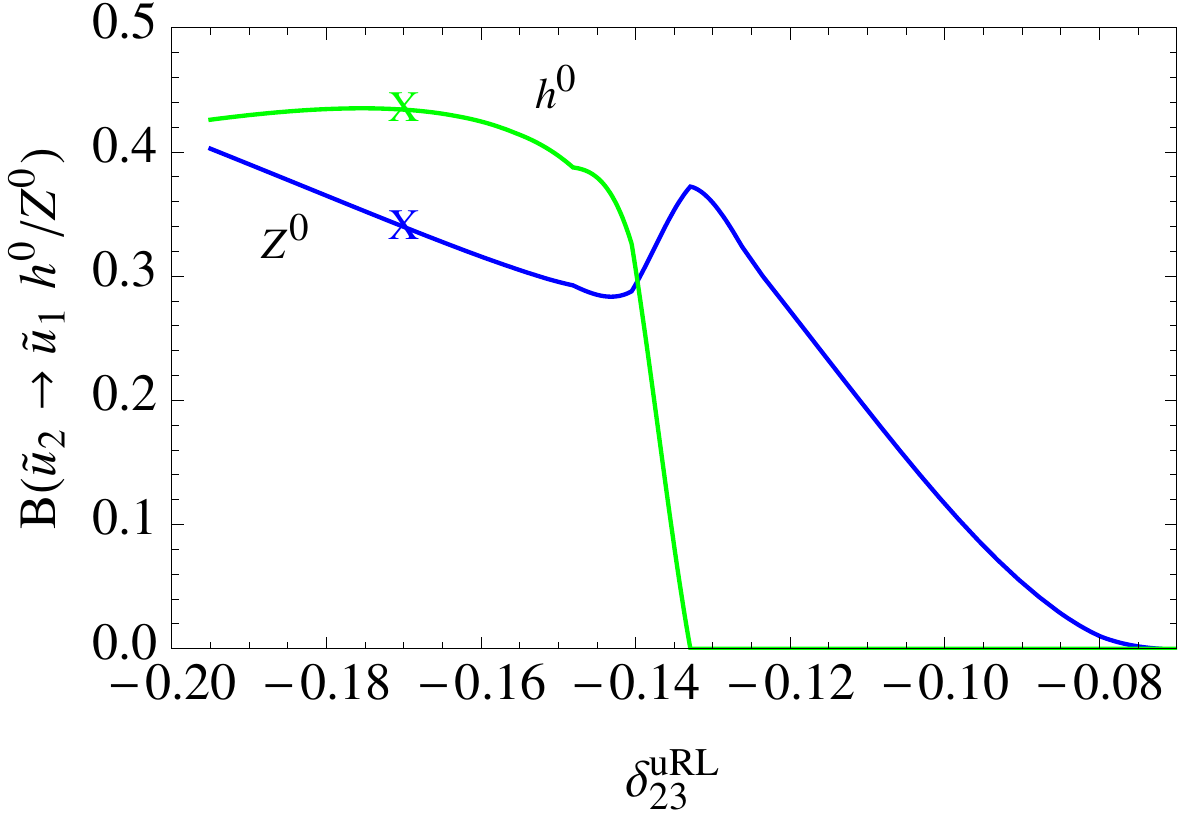}} \hspace*{-0.8cm}}}
\caption{$\durl$ dependence of the branching ratios B$(\su_2 \to \su_1 h^0)$ and B$(\su_2 \to \su_1 Z^0$) in scenario C. 
"X" indicates the reference point defined with 
the parameters of Table~\ref{basicparam}, except for those shown in 
eq.~(\ref{scenC}). 
The vanishing of  B$(\su_2 \to \su_1 h^0)$ at $\durl \approx -0.13$ is due to kinematics, which also causes the peak of   B$(\su_2 \to \su_1 Z^0$).}
\label{BRsu2tosu1h0Z0mod4a-3}
\end{figure}

In particular, the QFV trilinear coupling parameter $\durl (\sim T_{U32})$ is much larger than in scenario A.
This new scenario satisfies all experimental and theoretical 
constraints listed in Appendix~\ref{sec:exp_constr}. 
The physical masses, the flavour decomposition of $\su_1$ and $\su_2$ as well as the branching ratios of the two-body decays of the squarks and gluino in scenario C are shown in Table~\ref{physmassesC}, Table~\ref{flavourdecompC} and Table~\ref{BRsC}, respectively. 
As both B$(\su_2 \to \su_1 h^0)$ and B$(\su_2 \to \su_1 Z^0)$ 
are very large, this leads to the dominance of the QFV 
bosonic decays of $\su_2$. Note also that $\su_{1,2}$ are mixtures of 
$\sca_R$ and $\sto_L$ due to the sizable QFV trilinear coupling $T_{U32}$
$\sim \durl$, which 
significantly enhances the QFV decay $\su_2 \to \su_1 h^0$. 
The large B$(\su_2 \to \su_1 Z^0)$ is mainly due to the sizable $\sto_L$ 
component in $\su_{1,2}$ in this scenario, whereas in scenarios A and B
the $\sto_L$ component is small.
In Fig.~\ref{BRsu2tosu1h0Z0mod4a-3} we show B$(\su_2 \to \su_1 Z^0)$ and 
B$(\su_2 \to \su_1 h^0)$ as functions of $\durl$. We see that the ratio 
B$(\su_2 \to \su_1 Z^0)/$B$(\su_2 \to \su_1 h^0)$ is sensitive to the QFV
as well 
as to the QFC SUSY parameters. 
These results do not change for $\durl \to -\durl$.
%

\section{Characteristic final states}
\label{sec:signatures}

In this section we discuss some characteristic final states to be expected at LHC, $\sqrt{s}=14$~TeV, 
from the QFV decays of $\su_2$ into the lightest Higgs boson $h^0$ within the scenarios considered.
The lighter squark states can be produced directly, $pp \to \su_1 \bar{\su}_1 X$, $pp \to  \su_2 \bar{\su}_2 X$, or via gluino production, 
$pp \to \sg \sg X$, where at least one of the gluinos decays into $\su_1$ or $\su_2$, $\sg \to \su_{1,2}~ c; \su_{1,2}~ t$.
The $\su_{1,2}$ and gluino 
decays relevant for our study are as follows: 
\begin{eqnarray}
&& \su_1 \to c/ t ~\nt_1,  \\                                             
&& \su_2 \to c/ t ~\nt_1, \\                                                    
&& \su_2 \to \su_1 ~h^0/ Z^0 \to c/ t ~\nt_1 ~h^0/ Z^0,  \\                          
&& \sg \to \su_1 ~\bar{c}/ \bar{t} \to c/ t ~\nt_1~ \bar{c}/ \bar{t}~
({\rm and}~ c.c.), \\
&& \sg \to \su_2 ~\bar{c}/ \bar{t} \to c/ t ~\nt_1~ \bar{c}/ \bar{t} ~
({\rm and}~ c.c), \\
&& \sg \to \su_2 ~\bar{c}/ \bar{t} \to \su_1 ~h^0/ Z^0 ~\bar{c}/ \bar{t} \to  
c/ t ~\nt_1~ h^0/ Z^0 ~\bar{c}/ \bar{t}~ ({\rm and}~ c.c.). 
\end{eqnarray}         
We assume that $\nt_1$ is the lightest supersymmetric particle (LSP) and gives rise to missing transverse energy $\etmiss$ in experiment. The corresponding combined
decay 
branching ratios are given, for example, by 
\begin{eqnarray}
&& {\rm B}(\su_2 \to c/ t ~\nt_1~ h^0/ Z^0 ) =  {\rm B}(\su_2 \to \su_1
~h^0/ Z^0)
{\rm B}(\su_1 \to c/ t ~\nt_1), \\
&&  {\rm B}(\sg \to c/ t ~\nt_1 ~\bar{c}/ \bar{t}) = 2[{\rm B}(\sg \to
\su_1 ~\bar{c}/ \bar{t}) {\rm B}(\su_1 \to c/ t ~\nt_1)  \nonumber \\
&& \hspace*{5cm}
+{\rm B}(\sg \to \su_2 ~\bar{c}/ \bar{t}){\rm B}(\su_2
\to c/ t ~\nt_1)], \\
&& {\rm B}(\sg \to c/ t ~\nt_1 ~h^0/ Z^0 ~\bar{c}/ \bar{t}) = 2 {\rm
B}(\sg \to \su_2 ~\bar{c}/ \bar{t})  \nonumber \\
&& \hspace*{5cm} \times
{\rm B}(\su_2 \to \su_1 ~h^0/ Z^0) {\rm
B}(\su_1 \to c/ t ~\nt_1).
\end{eqnarray}   
In Table~\ref{tab:signatures} we list the processes leading to at least one Higgs boson $h^0$ 
in the final state in association with jets and top-quarks. We assume that the 
$c$-quarks hadronize to jets, similarly to $u$-quarks. Of course, additional $c$-tagging would be very helpful. 
\begin{table}[h!]
\caption{
Possible final states containing at least one Higgs boson $h^0$ 
expected from the decays of $\su_2$ into $h^0$ and $Z^0$. 
$t$ denotes top-quark or anti-top-quark; $j$ denotes a $c/\bar{c}$-quark jet; 
$\etmiss$ is missing transverse energy due to the two LSP neutralinos $\nt_1$
in the final state; $X$ contains only the beam jets. Note that in general the states 
with $h^0$ replaced by $Z^0$ are also possible.
We also give the corresponding cross sections
in scenario A, 
in case they exceed 1~fb.  
We indicate by "QFV" the final states which are explicitely QFV. } 
\label{tab:signatures}
\begin{center}
\begin{tabular}{|c|cc|}
\hline
 processes & final states  containing $h^0$ &\\ 
\hline & &\\
 $pp \to \su_2 \bar{\su}_2 X$ & $2 j +h^0 + \etmiss+X$ (1.5~fb)&\\
$$ & $j+ t + h^0 + \etmiss+X$ (2.8~fb);& QFV\\
 $$ & $2 t +     h^0 + \etmiss+X$ &\\
 $$ & $2 j +   2 h^0 + \etmiss+X$ &\\
 $$ & $j+ t +  2 h^0 + \etmiss+X$ (1~fb); & QFV\\
 $$ & $2 t +   2 h^0 + \etmiss+X$ &\\
$$ & $2 j +     h^0 + Z^0 + \etmiss+X$ &\\
 $$ & $ j + t +   h^0 + Z^0 + \etmiss+X$; & QFV\\  
 $$ & $ 2 t +     h^0 + Z^0 + \etmiss+X$ &\\ 
 &&\\
 %
\hline
\end{tabular}
\vskip 0.4cm
\begin{tabular}{|c|cc|}
\hline
processes & final states containing $h^0$ &\\ 
 \hline &&\\
 $pp \to \sg \sg X$ & $4j+        h^0 + \etmiss+X$ (2~fb) &\\
& $3j+ t +  h^0 + \etmiss+X$ (8~fb); & QFV\\
& $2j+ 2t + h^0 + \etmiss+X$ (13~fb); & 8~fb QFV\\ 
& $4j+ 2h^0 + \etmiss+X$ &\\ 
& $3j+ t+ 2h^0 + \etmiss+X$; & QFV\\ 
& $2j+ 2t + 2h^0 + \etmiss+X$ &\\ 
& $4j+ h^0 + Z^0 + \etmiss+X$ &\\ 
& $3j+ t + h^0 + Z^0 + \etmiss+X$; & QFV\\ 
& $2j+ 2t + h^0 + Z^0 + \etmiss+X$ &\\
& &\\
  \hline
\end{tabular}
\end{center}
\end{table}

In Table~\ref{tab:signatures}, $t$ denotes a top-quark or an anti-top-quark and 
$j$ denotes a $c/\bar{c}$-quark jet. 
What concerns the final states from $\su_2 \bar{\su}_2$ pair production, the final states with 
one $t$ are explicitly QFV whereas those with no $t$ and 2$t$
look like QFC. 
The cross sections for $pp \to \su_2 \su_2 X$ and $pp \to \bar{\su}_2 \bar{\su}_2 X$ are smaller than 1 fb.
Concerning the final states from gluino pair production, 
those with one $t$ and 3$t$ are explicitly QFV whereas 
those with no $t$, 2$t$ and 4$t$ look like QFC.
Note that the final states with 3t and 4t are not shown in Table~\ref{tab:signatures} since 
the corresponding cross sections are very small (much less than 1 fb). 
The states with $2t$ are explicitly QFV in case they are $tt$ or $\bar{t} \bar{t}$. On the other hand, they look like QFC in case they are $t \bar{t}$. The events with $t \bar{t}$ can stem from QFV and QFC gluino decays, e.g.
$\sg \sg \to (c \bar{t} h^0 \nt_1) + (t \bar{c} h^0 \nt_1)$ and 
$\sg \sg \to (t \bar{t} h^0 \nt_1) + (c \bar{c} h^0 \nt_1)$. 
Note also that the events with $t t ($or$~\bar{t}
\bar{t}) jj$, such as $t t$ (or $\bar{t}
\bar{t}) jj h^0 \etmiss X$ 
(where $X$ contains 
only the beam jets)
can practically not 
be produced in the QFC MSSM (nor in the SM). 
The
detection of such events could be useful for 
discriminating between the QFC MSSM and QFV MSSM. 

For scenario A, the production cross section for $pp \to \sg \sg X$ is 148~fb
including one-loop SUSY-QCD corrections. For
calculating this cross section
we have used Prospino 2~\cite{sgsgNLO_Prospino}
as the cross section is only very weakly dependent on the QFV parameters.
The cross sections for $pp \to \su_2 \bar{\su}_2 X$,  $pp \to \sg \su_1 X$, and $pp \to \sg \su_2 X$ are 
at tree-level 10~fb, 1~fb, and 1.4~fb, respectively.   
For the calculation of these cross sections we have used FeynArts and FormCalc~\cite{FAref,FAFCref}.
All numbers for the cross sections given in this section include the charge conjugate final states.

In scenario A (Table~\ref{basicparam}), using the decay branching ratios of $\su_2$ and $\su_1$, as shown in Table~\ref{BRsA}, 
we find that the produced $\su_2 \bar{\su}_2$ state goes into the final 
state $2j +h^0+\etmiss$ with a probability of 15\%.
Hence, in our scenario, the corresponding cross section for $pp \to \su_2 \bar{\su}_2 X \to 2j+h^0+\etmiss+X$ is about 1.5~fb. Note, however, that this final state can also occur in the QFC bosonic decays. 
On the other hand, the process $pp \to \su_2 \bar{\su}_2 X \to j + t + h^0 
+ \etmiss+X$ is QFV and the corresponding cross section is almost 2.8~fb.
Even the cross section for $pp \to \su_2 \bar{\su}_2 X \to j + t + 2h^0
+\etmiss+X$ 
is about 1~fb.
As the ratio B$(\su_2 \to \su_1 Z^0)$/ B$(\su_2 \to \su_1 h^0)$ is only about 0.02 in scenario A,
the probability for the $\su_2 \bar{\su}_2$ system to decay into the final
state 
$2j+Z^0+h^0+\etmiss$ is only 0.1$\%$.
%
\begin{figure*}[h!] 
\centering
\subfigure[]{
   { \mbox{\hspace*{-1cm} \resizebox{8.cm}{!}{\includegraphics{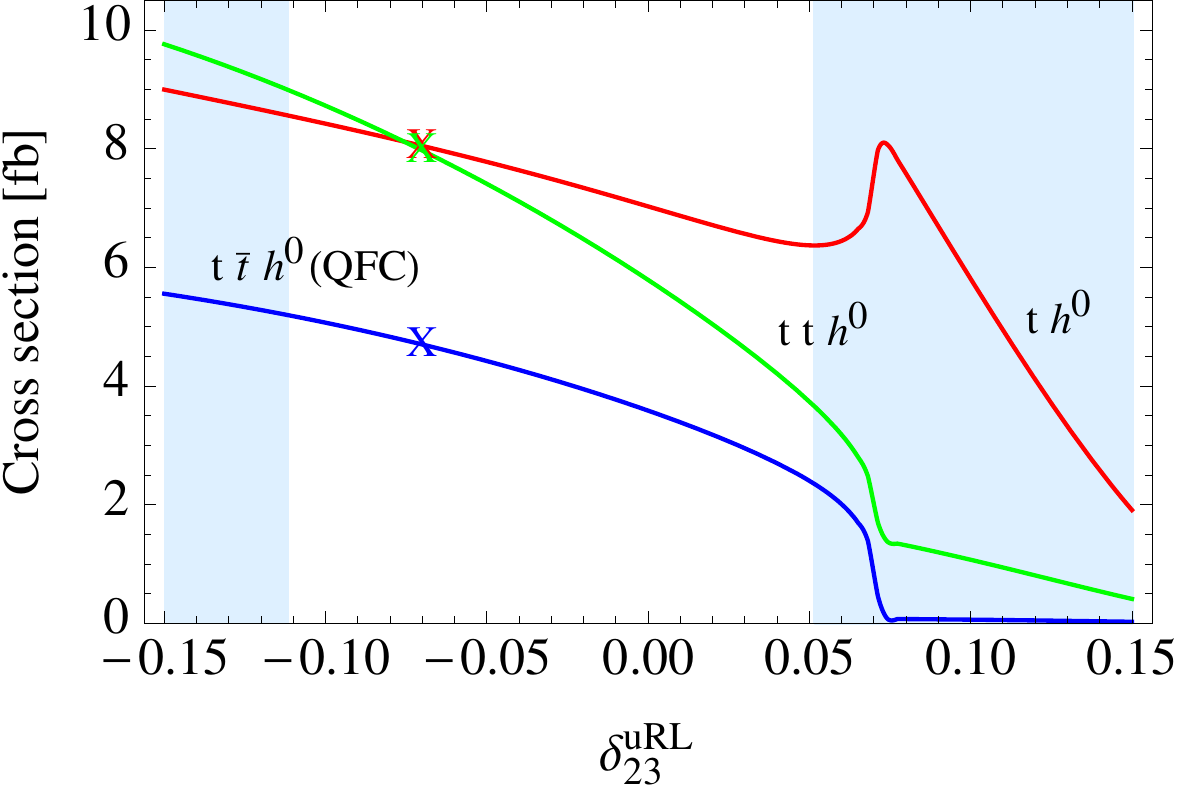}} \hspace*{-0.8cm}}}
  \label{signature_scenA}}
 \subfigure[]{
   { \mbox{\hspace*{+0.cm} \resizebox{8.cm}{!}{\includegraphics{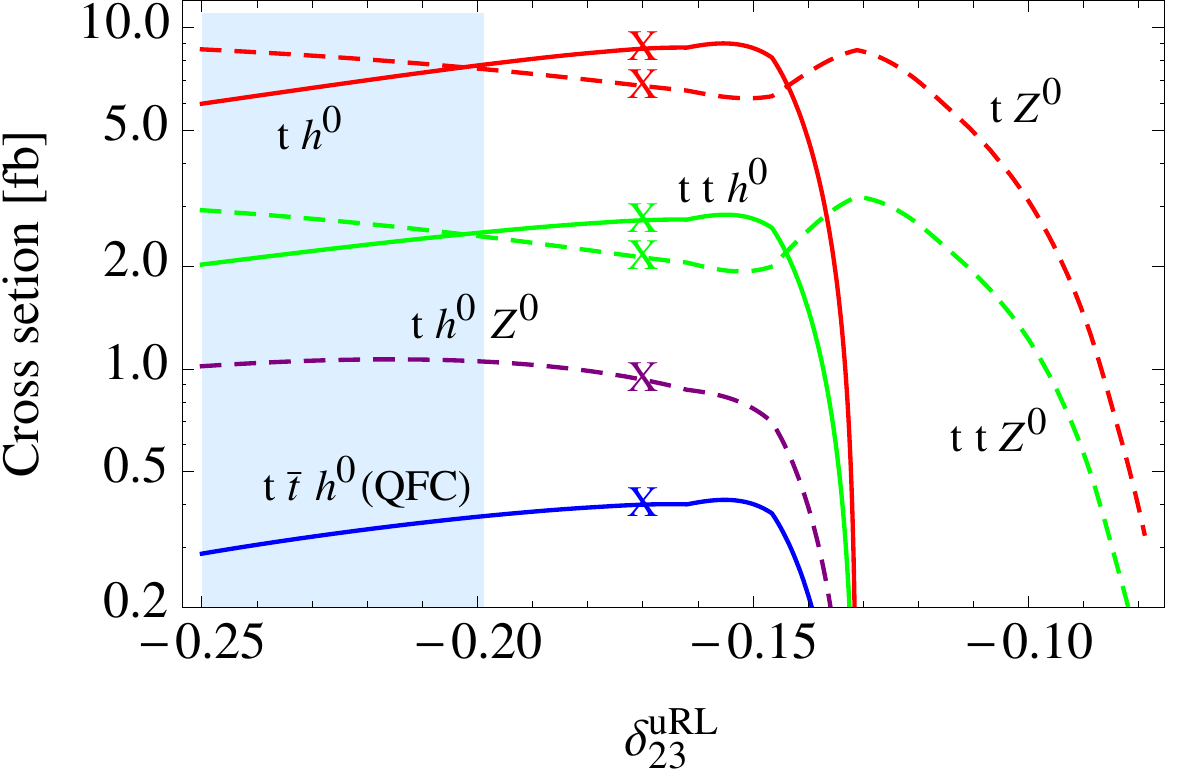}} \hspace*{-1cm}}}
  \label{signature_scenC}}
\caption{Cross sections for the final states coming
from gluino pair production and subsequent decays $\sg \to \su_{1,2}~ c/t,~\su_2 \to \su_1 h^0/Z^0$ at LHC, $\sqrt{s}=14$~TeV, (a) 
in scenario A and (b) in scenario C as functions of $\durl$.
The red solid (dashed) line corresponds to the pure QFV final state $3j+t+h^0~(Z^0)+\etmiss+X$.
The green solid (dashed) line corresponds to the QFV events $tt/  \bar{t}\bar{t}/ t\bar{t}+ 2j+h^0~(Z^0)+\etmiss+X$ coming from the QFV
gluino decays. 
The blue solid line corresponds to the events $t+\bar{t}+2j+h^0+\etmiss+X$ coming from the QFC
gluino decays.
The violet dashed line corresponds 
to the pure QFV final state $3j+t+ h^0 + Z^0 +\etmiss+X$.
"X" indicates the corresponding scenario's reference point: for scenario A defined with the parameters of Table~\ref{basicparam}, and for scenario C defined with the parameters of Table~\ref{basicparam}, except for those shown in eq.~(\ref{scenC}). The shaded (light blue) 
areas are excluded by $m_h^0<123~\gev$.
 \label{signatures_scenAC}}
\end{figure*}

The squarks $\su_2$ are also produced in gluino decays. 
The branching ratios for the gluino decays at the reference point of scenario A are given in Table~\ref{BRsA}.
For the process $\sg \sg \to \su_2 \su_2 jj \to 4j+h^0+\etmiss+X$ one gets a probability of 0.8$\%$, leading to a cross section for $pp \to \sg \sg X \to \su_2 \su_2 jj X \to 4j+h^0+\etmiss+X$ of about 1.2~fb.
(Here the contributions of $pp \to \sg \sg X \to \su_2 \bar{\su}_2 jj X$ are also included.)
In the process $pp \to \sg \sg X \to \su_2 \su_2 jj X$ the final state $4j+2h^0+\etmiss+X$ is possible with a probability of approximately 0.3$\%$. The cross section for $pp \to \sg \sg X \to 4j+2h^0+\etmiss+X$ is 0.5~fb.

A further interesting process is $pp \to \sg \sg X \to \su_1 \su_2$ $ jj X$ $\to 4j+h^0+\etmiss+X$, having a probability of 0.5$\%$, 
giving a cross section of 0.8~fb. 
(We have included also $pp \to \sg \sg X \to \su_1 \bar{\su}_2$.)
Therefore, one has a cross section of 2~fb altogether for $pp \to \sg \sg X \to 4j+h^0+\etmiss+X$. The QFV 
final state $3j+t+h^0+\etmiss+X$ coming from $pp \to \sg \sg X$ has a cross section of 8~fb.  Correspondingly, 
the final state $2j+2t+h^0+\etmiss+X$ from  $pp \to \sg \sg X$ has a cross section of 13~fb, containing
a QFC contribution of 5~fb (see also Fig.~\ref{signature_scenA}). The cross section of 
$pp \to \sg \sg X \to 3j+t+2h^0+\etmiss+X$ is almost 0.9~fb. 
In Table~\ref{tab:signatures} we give the corresponding cross sectios in case they exceed 1~fb. 

Summing up the cross sections for all final states with at least one $h^0$ 
in scenario A one gets 28~fb, 16~fb of which come from pure QFV final states. This means that one could expect about 
1600 of such events assuming an integrated luminosity of 100~\fb at LHC (14 TeV).

In Fig.~\ref{signature_scenA} we show the cross sections for $pp \to \sg \sg X \to 3j+t+h^0+\etmiss+X$ and 
$pp \to \sg \sg X \to 2j+2t+h^0+\etmiss+X$
in scenario A as a function of $\durl$. 
The red solid line corresponds to the pure QFV final state $3j+t+h^0+\etmiss+X$. 
The green solid line corresponds to the pure QFV final state $tt/  \bar{t}\bar{t}+ 2j+h^0+\etmiss+X$ plus the 
final state $t\bar{t}+ 2j+h^0+\etmiss+X$ coming from the QFV gluino decays. 
Note that the 
number of the 
$tt/\bar{t}\bar{t}$ final state events is exactly equal to the
number of the 
$t\bar{t}$ final state events coming from the QFV gluino decays due to the Majorana nature of the gluino.
The blue solid line corresponds to the QFC events $t+\bar{t}+2j+h^0+\etmiss+X$ coming from the QFC gluino decays.
At the reference point the QFV and QFC cross sections for the final states containing $t t h^0$ are 8~fb and 5~fb, respectively (see also Table~\ref{tab:signatures}).
For $\durl > 0.074$ the decay $\su_1 \to t \nt_1$ is kinematically not possible whereas the decay $\su_2 \to t \nt_1$ is still allowed (with 18$\%$ branching ratio). 
Note that the QFV cross sections do not vanish for $\durl=0$ because 
the other QFV parameter $\delta^{uRR}_{23}$ is not zero.

In the GUT inspired scenario (scenario B) the gluino is much heavier and therefore the cross section $\sigma(pp \to \sg \sg X)$
is much smaller being 3.5~fb. The final state coming from $pp \to \su_2 \bar{\su}_2 X$ have the same cross section as in scenario A, whereas the cross sections for the final states due to $pp \to \sg \sg X$ are about a factor of 40 smaller.

The third scenario (scenario C) is characterized by a higher branching ratio B($\su_2 \to \su_1 Z^0)=34\%$, see Table~\ref{BRsC}. Therefore, one expects final states with $Z^0$ and $h^0$. 
As the gluino mass is very close to that of scenario A, the cross section for $pp \to \sg \sg X$ is 148~fb.
Consequently, the cross section $\sigma(pp \to \sg \sg X \to 3j + t + h^0 + \etmiss+X)$ is 8.5~fb and  $\sigma(pp \to \sg \sg X \to 3j + t + Z^0 + \etmiss+X)$ is 6.8~fb.
In Fig.~\ref{signature_scenC} we show the cross sections analogous to those shown in Fig.~\ref{signature_scenA}, 
but for scenario C. In addition, we also show the cross sections of the final states containing a $Z^0$.
The green dashed line corresponds to the pure QFV final state $tt/  \bar{t}\bar{t}+ 2j+Z^0+\etmiss+X$ plus the 
final state $t\bar{t}+ 2j+Z^0+\etmiss+X$ coming from the QFV gluino decays. The number of the 
$tt/\bar{t}\bar{t}$ final state events is again equal to the number of the QFV $t\bar{t}$ final state events.
The violet dashed line corresponds to the pure QFV final state $3j+t+h^0+Z^0+\etmiss+X$. 
 Fig.~\ref{signature_scenC} is symmetric for $\durl \to -\durl$.

We want to comment shortly on the background processes to the QFV bosonic
squark 
decay signals containing at least one Higgs boson $h^0$. 
An important background is the production of a Higgs boson $h^0$ in association with top quarks, 
$pp \to t \bar{t} h^0 X$, where $h^0$ is radiated off from top or anti-top. The cross section at $\sqrt{s}=14$~TeV 
is about 400~fb. In these events, however, there is no missing energy, $\etmiss$ (apart from the missing energy coming from possible semi-leptonic decays of the
top-quarks), therefore it should be possible to separate 
them from the signal. Further Higgs 
boson production processes are $pp \to Z^0Z^0h^0; W^+W^-h^0$. They will of course, constitute a 
background to the $h^0+jets+\etmiss$. However, these processes do not contain a top in the final state. 
As discussed above, events with top (anti-top) in the final states together with a $h^0$ are the most significant 
ones for QFV. Single $h^0$ production from gluon-gluon fusion as well as $pp \to b \bar{b} h^0 X$ do not 
contain a top quark in the final state either.

Concerning the background within the general MSSM, the situation can be more complex. In the scenarios considered 
the charginos and neutralinos are relatively heavy, so that the decays of the lightest squarks $\su_{1,2}$ into these 
play a minor role, except those into the lightest neutralino. If this is not the case the QFV signals will be less pronounced. 

The most interesting final states exhibiting QFV in bosonic squark decays are $j+t+h^0+\etmiss+X$ 
from $\su_2 \bar{\su}_2$ production and $3j+t+h^0+\etmiss+X$ from $\sg \sg$ production. To extract these events, the 
identification of the top-quark and the Higgs boson by their decay products would be crucial. This would require Monte 
Carlo studies including appropriate cuts and detector simulation. However, this is beyond the scope of this paper.

\section{Summary}
\label{sec:conclude}

In this paper we have studied the effects of QFV in the bosonic squark 
decays $\su_2 \to \su_1 h^0/Z^0$
at the LHC. We have assumed mixing between the second and third up-squark generations, 
that is $\sca_{R}-\sto_{L,R}$ mixing. 
In our calculations, we have taken into 
account all experimental constraints
from B meson data on $\Delta M_{B_s}$, B$(b \to s \gamma)$, B$(B_s \to 
\mu \mu)$, limits on the gluino and squark masses, the latest data on the 
lightest Higgs boson mass
and the theoretical constraints on the 
trilinear couplings from the vacuum stability conditions. We have found that 
the branching ratio B$(\su_2 \to \su_1 h^0)$ can be larger than in the 
QFC case, and can go up to $50\%$.
The decay $\su_2 \to \su_1 
h^0$ can give access to the QFV trilinear couplings $T_{U32}$ and $T_{U23}$. 
We have studied the characteristic final states expected from the QFV decay $\su_2 
\to \su_1 h^0$ at LHC with $\sqrt{s}=14$~TeV in three different 
scenarios. We have considered direct $\su_2$ production $pp \to \su_2 
\bar{\su}_2 X$ as well as $\su_2$ production in $\sg$ decays via $pp 
\to \sg \sg X$. 
In two scenarios (A and C) we have taken $\msg \approx 1100~\gev$ and in the third  scenario (B)  $\msg \approx 1600~\gev$.
The most pronounced 
QFV final state is $3j+t+h^0+\etmiss+X$, coming from 
$pp \to \sg \sg X \to \su_{1,2}\bar{t}  \su_2 \bar{c} X \to \su_{1,2}\bar{t} \su_1
h^0 \bar{c}
X \to c \bar{t} c \bar{c} h^0 \etmiss X$, which can have a cross section up 
to 8~fb in scenario A.
For extracting these events, an identification of the top quark and the Higgs boson would be required.

In conclusion, our analysis suggests that for a complete determination 
of the parameters of the squark mass matrices in the MSSM it would be necessary to study both the fermionic 
and the bosonic QFC and QFV decays of squarks. This can also have an 
influence on the squark and gluino searches at LHC. 
%
\section*{Acknowledgments}

This work is supported by the "Fonds zur F\"orderung der
wissenschaftlichen Forschung (FWF)" of Austria, project
No. I 297-N16 and project No. P26338-N27,
by the DFG, project No.
PO-1337/2-1 and by DAAD, project PROCOPE 54366394.

%
\begin{appendix}


\section{Up-squark decays into \boldmath{$h^0$}}
\label{sec:sq.coupl}

\begin{table*}[h!]
\footnotesize{
\caption{
Constraints on the MSSM parameters from the B-physics experiments
relevant mainly for the mixing between the second and the third
generations of
squarks and from the limit on the $h^0$ mass. The fourth column shows
constraints 
at $95 \%$ CL obtained by combining the experimental error quadratically
with the
theoretical uncertainty, except for $m_{h^0}$.
$R^{\rm SUSY}_{B \to \tau \nu}=\frac
{ {\cal B}^{\rm {SUSY}} (B_u \to \tau \nu)}
{ {\cal B}^{\rm {SM}} (B_u \to \tau \nu)}
= [1-(\frac{m_B^2}{m_{H^\pm}^2})\frac{\tan^2\beta} {(1+\epsilon_0 \tan \beta)}]^2$,
where $|\epsilon_0| \lesssim 10^{-2}$ and $ m_{H^\pm}$ is the $H^\pm$ mass ~\cite{Isidori:2006pk, Hou:1992sy}.
}
\begin{center}
\begin{tabular}{|c|c|c|c|}
    \hline
    Observable & Exp.\ data & Theor.\ uncertainty & \ Constr.\ (95$\%$CL) \\
    \hline\hline
    &&&\\
    $\Delta M_{B_s}$ [ps$^{-1}$] & $17.725 \pm 0.049$ (68$\%$ CL)~\cite{
    LHCb2011@ICHEP2012} & $\pm 3.3$ (95$\%$ CL)~\cite{DeltaMBs_R22,Ball:2006xx} &
$17.73 \pm 3.30$\\
    $10^4\times$B($b \to s \gamma)$ & $3.37 \pm 0.23$ (68$\%$ CL)~\cite{
    Stone@ICHEP2012} & $\pm 0.23$ (68$\%$ CL)~\cite{BsgTheo2} &  $3.37\pm
    0.64$\\
    $10^6\times$B($b \to s~l^+ l^-$)& $1.60 \pm 0.50$ (68$\%$ CL)~\cite{Bsll_R5,Aubert:2004it}&
$\pm 0.11$
    (68$\%$ CL)~\cite{Huber:2007vv}& $1.60 \pm 1.00$\\
    $(l=e~{\rm or}~\mu)$ &&&\\
$10^9\times$B($B_s\to \mu^+\mu^-$) & $ < 4.2 $ (95$\%$ CL)~\cite{Bstomumu} 
&  & $1.4 < 10^9 \times B < 4.2$ \\
                      & $2.9 \pm 0.7$ (68$\%$CL)~\cite{LHCb_Bsmumu_evidence}
& $\pm0.27$  (68$\%$ CL)~\cite{Buras,Buras:2013uqa}&  \\

$10^4\times$B($B^+ \to \tau^+ \nu $) & $1.15 \pm 0.23$  (68$\%$
CL)~\cite{Btotaunu,Lees:2012ju,Adachi:2012mm} 
&$\pm0.29$  (68$\%$ CL)~\cite{Btotaunu,Lees:2012ju,Adachi:2012mm} & $1.15 \pm 0.73$\\
&&&$R^{\rm SUSY}_{B \to \tau \nu}=$\\
&&&$1.14\pm0.78$\\
$ m_{h^0}$ [GeV] & $125.3 \pm 0.64~(68\%~ \rm{CL}) (\rm{CMS}), $
&&\\
&$126.0 \pm 0.57~(68\%~ \rm{CL})(\rm{ATLAS})$
 & $\pm 2$~\cite{Heinemeyer:2011aa} & $123 < m_{h^0} < 129 $\\
&\cite{CMS@ICHEP2012, ATLAS@ICHEP2012,ATLAS:
2012ae,Chatrchyan:2012tx,Chatrchyan:2012ufa}&&\\
&&&\\
    \hline
\end{tabular}
\end{center}
\label{TabConstraints}}
\end{table*}
%

In the super-CKM basis, the Lagrangian including the coupling of up-type squarks to the 
lighter neutral Higgs boson, $h^0$, is given by
\begin{eqnarray} 
&&{\cal L} = - \frac{g_2}{2 m_W}\,h^0 \Bigg[\
 \su^*_{ i L} \su_{j L} \Big(-
m_W^2 \sin(\alpha+\beta)
(1-\frac{1}{3} \tan^2\theta_W)\ \delta_{ij}
+ 2\, \frac{\cos\alpha}{\sin\beta}\ m^2_{u,i}\ \delta_{ij} \Big) \nonumber \\
&& +\, \su^*_{iR} \su_{jR} \Big(
-m_W^2 \sin(\alpha+\beta) \frac{4}{3} \tan^2\theta_W \delta_{ij} 
+ 2\, \frac{\cos\alpha}{\sin\beta} m^2_{u,i} \delta_{ij} \Big)\nonumber \\
&&+\ \Big[\   \su^*_{iR} \su_{jL} \Big(
 \mu^* \frac{\sin\alpha}{\sin\beta} m_{u,i} \delta_{ij} 
+ \frac{\cos\alpha}{\sin\beta} \frac{v_2}{\sqrt2} (T_U)_{ji}
 \Big) + \rm {h. c.} \Big] \Bigg] \,, \label{eqSuSuh0}  \nonumber \\
\end{eqnarray}  
where $\alpha$ is the mixing angle of the two CP-even Higgs bosons, $h^0$ and $H^0$.
The terms proportional to $m^2_W$ stem from the D-terms of the scalar potential and the expressions with quark masses $m_{u,d}$ stem from Yukawa and F-terms. They are all flavour-universal. The trilinear couplings are explicit breaking terms that couple left-handed to right-handed squarks. 
Inserting the transformations to the physical fields, $\su_{iL} = (R^{\su \dagger})_{ik} \su_k$
and $\su_{iR} = (R^{\su \dagger})_{(i + 3)\,k} \su_k$, eq.~(\ref{eqSuSuh0}) can 
be written in terms of physical up-type squark fields as 
${\cal L} = c_{\su_i \su_j h^0}\, \tilde{u}^*_j\, \tilde u_i \, h^0$ with the coupling
\begin{eqnarray}
&&  c_{\su_i \su_j h^0} =  -\frac{g_2}{2 m_W}\,
	\bigg[-m_W^2 \sin(\alpha+\beta) \Big[
      (1-\tfrac13 \tan^2\thw) \nonumber \\
    &&  \times (R^{\su})_{jk} (R^{\su \dagger})_{ki} + \tfrac43 \tan^2\thw (R^{\su})_{j\,(k+3)} 
    (R^{\su \dagger})_{(k+3)\,i}
      \Big] \nonumber \\[0.2cm]
    & & +\ 2 \dfrac{\cos\alpha}{\sin\beta} \Big[
      (R^{\su})_{jk}\ m^2_{u,k} (R^{\su \dagger})_{ki} 
      + (R^{\su})_{j\,(k+3)} m^2_{u,k} (R^{\su \dagger})_{(k+3)\,i}
      \Big] \nonumber \\[0.2cm]
    & & +\ \dfrac{\sin\alpha}{\sin\beta} \Big[
      \mu^* (R^{\su})_{j\,(k+3)} m_{u,k} (R^{\su \dagger})_{ki} 
      + \mu (R^{\su})_{jk} m_{u,k} (R^{\su \dagger})_{(k+3)\,i}
      \Big] \nonumber \\[0.2cm]
    & & +\ \dfrac{\cos\alpha}{\sin\beta}\, \dfrac{v_2}{\sqrt2} \Big[
      (R^{\su})_{j(k+3)}\ (T_U)_{lk}\ (R^{\su \dagger})_{li} 
      + (R^{\su})_{jk}\ (T_U^\dagger)_{lk}\ (R^{\su \dagger})_{(l+3)\,i}
    \Big] \bigg] \, ,
    \label{coupluiujh}
\end{eqnarray}
where the sum over $k,l=1,2,3$ is understood. 
The decay width for the process $\su_i \to \su_j h^0$ is given by
\begin{equation}
\Gamma(\su_i \to \su_j h^0) = \frac{1} {16 \pi}\, \frac{\kappa(m^2_{\su_i}, m^2_{\su_j}, h^0)} {2 m^3_{\su_i}}\,
| c_{\su_i \su_j h^0} |^2\,.
\label{decratesuisujh0}
\end{equation}
As usual, $\kappa$ is defined by  $\kappa^2( x, y, z) = (x - y - z)^2 - 4 y z$.

\section{Experimental and theoretical constraints}
\label{sec:exp_constr}

Here we summarize the experimental and theoretical constraints taken into 
account in the present paper. The constraints on the MSSM parameters from the 
B-physics experiments and from the Higgs boson search at LHC are shown in 
Table~\ref{TabConstraints}. 
Recently the BaBar collaboration has reported a slight 
excess of B$(B\to D\, \tau\,\nu)$ and B$(B\to D^*\,\tau\,\nu)$ \cite{Lees:2012xj, Lees:2013uzd}.
However, it has been argued in \cite{Crivellin:2012ye} that within the
MSSM this cannot be explained without being at the same time in conflict
with B$(B_u\to \tau\,\nu)$. 
Using the program SUSY\_FLAVOR \cite{Crivellin:2012jv} we have checked that 
in our MSSM scenarios no significant enhancement occurs for B$(B\to D\, \tau\,\nu)$.
However, as pointed out in \cite{Nierste:2008qe}, the theoretical 
predictions (in SM and MSSM) on B$(B \to D\, l\, \nu)$ and B$(B \to D^*\, l\, \nu)$~$(l = \tau, 
\mu, e)$ have potentially large theoretical uncertainties due to the theoretical 
assumptions on the form factors at the $B\,D\,W^+$ and $B\,D^*\,W^+$ vertices 
(also at the  $B\,D\,H^+$  and $B\,D^*\,H^+$vertices in the MSSM). Hence the 
constraints from these decays are unclear. Therefore, we do not take these 
constraints into account in our paper.

The particle discovered most recently at LHC~\cite{CMS@ICHEP2012, ATLAS@ICHEP2012,
ATLAS:2012ae, Aad:2012tfa,Chatrchyan:2012tx,Chatrchyan:2012ufa} is consistent with the SM Higgs boson.  
We identify this particle as the MSSM Higgs boson $h^0$ which is indeed 
SM-like in the decoupling Higgs scenarios considered in our paper.
For the mass of the Higgs boson $h^0$, we take an average of the central 
values of the ATLAS and CMS data~\cite{ATLAS:2012ae, Aad:2012tfa, Chatrchyan:2012tx,Chatrchyan:2012ufa} 
and adding the theoretical uncertainty of $\sim \pm 2~\gev$
~\cite{Heinemeyer:2011aa} linearly to the experimental 
uncertainty at 2 $\sigma$, we take $123~\gev < m_{h^0} < 129~\gev$.

In addition to these constraints we also require that our scenarios are 
consistent with the following experimental constraints: 

(i) The LHC limits on the squark and gluino masses (at 95\% CL) 
    ~\cite{SUSY@ICHEP2012, SUSY2@ICHEP2012,SUSY@LHC, Aad:2012ms, AtlasConf2012, Aad:2012hm, SUSY1@LHC, SUSY2@LHC, Chatrchyan:2012rg, Chatrchyan:2012ira, Chatrchyan:2012qka, stop_sbot@LHC, Aad:2012yr, Aad:2012tx, Aad:2012xqa, Aad:2012ywa, Aad:2012pq, ATLAS:2012ah, ATLAS:2012ai, Aad:2011cw, CMS,Chatrchyan:2013lya}: 
In the context of simplified models, gluino masses $\msg \lesssim 1~{\rm TeV}$ are excluded at 95\% CL. The mass limit varies in the range 950-1125~GeV. 
First and second generation squark masses are excluded below 775~GeV. Bottom squarks are excluded below 600~GeV.
In~\cite{CMS,Chatrchyan:2013lya} a limit for the mass of the top-squark $m_{\tilde{t}} \gsim 500~\gev$ for $m_{\tilde{t}} - m_{\rm LSP}=200~\gev$ is quoted. 
Including mixing of $\sca_R$ and $\sto_R$ would even lower this limit~\cite{Blanke}. 

(ii) The LHC limits on $m_{\ch_1}$ and $m_{\nt_1}$ from negative
searches for charginos and neutralinos in leptonic final states
~\cite{chargino_neutralino@LHC,Chatrchyan:2012pka}.

(iii) The constraint on ($m_{A^0}$ , $\tan\be$) from the MSSM Higgs boson
searches at LHC 
      ~\cite{MSSM_Higgs@LHC}.

(iv) The experimental limit on SUSY contributions on the electroweak
$\rho$ parameter
     ~\cite{Altarelli:1997et}: $\Delta \rho~ (\rm SUSY) < 0.0012.$

Furthermore, we impose the following theoretical constraints from the vacuum 
stability conditions for the trilinear coupling matrices~\cite{Casas}: 
\begin{eqnarray}
|T_{U\alpha\alpha}|^2 &<&
3~Y^2_{U\alpha}~(M^2_{Q \alpha\alpha}+M^2_{U\alpha\alpha}+m^2_2)~,
\label{eq:CCBfcU}\\[2mm]
|T_{D\alpha\alpha}|^2 &<&
3~Y^2_{D\alpha}~(M^2_{Q\alpha\alpha}+M^2_{D\alpha\alpha}+m^2_1)~,
\label{eq:CCBfcD}\\[2mm]
|T_{U\alpha\beta}|^2 &<&
Y^2_{U\gamma}~(M^2_{Q \alpha\alpha}+M^2_{U\beta\beta}+m^2_2)~,
\label{eq:CCBfvU}\\[2mm]
|T_{D\alpha\beta}|^2 &<&
Y^2_{D\gamma}~(M^2_{Q\alpha\alpha}+M^2_{D\beta\beta}+m^2_1)~,
\label{eq:CCBfvD}
\end{eqnarray}
where
$\al,\be=1,2,3,~\al\neq\be;~\gamma={\rm Max}(\al,\be)$ and
$m^2_1=(m^2_{H^\pm}+m^2_Z\sin^2\theta_W)\sin^2\beta-\frac{1}{2}m_Z^2$,
$m^2_2=(m^2_{H^\pm}+$  
$m^2_Z\sin^2\theta_W)$ $\cos^2\beta-\frac{1}{2}m_Z^2$.
The Yukawa couplings of the up-type and down-type quarks are
$Y_{U\alpha}=\sqrt{2}m_{u_\alpha}/v_2=\frac{g}{\sqrt{2}}\frac{m_{u_\alpha}}{m_W
\sin\beta}$
$(u_\al=u,c,t)$ and
$Y_{D\alpha}=\sqrt{2}m_{d_\alpha}/v_1=\frac{g}{\sqrt{2}}\frac{m_{d_\alpha}}{m_W
\cos\beta}$
$(d_\al=d,s,b)$,
with $m_{u_\al}$ and $m_{d_\al}$ being the running quark masses at the weak
scale and
$g$ being the SU(2) gauge coupling. All soft-SUSY-breaking parameters are
given at
$Q=1$~TeV. As SM parameters we take $m_W=80.4~\gev$,
$m_Z=91.2~\gev$ and
the on-shell top-quark mass $m_t=173.3~\gev$ \cite{Shabalina}.
We have found that our results shown are fairly
insensitive to the
precise value of $m_t$.

\end{appendix}

%
%

\end{document}